\documentclass[nofootinbib,amsmath,prd,aps,superscriptaddress,preprintnumbers,tightenlines,12pt]{revtex4}
\usepackage{graphicx}
\usepackage{epstopdf}
\usepackage{bm}
\usepackage{epsfig}
\usepackage{graph ics}
\usepackage{xspace}
\usepackage{color}
\usepackage{subfigure}

\usepackage{amsmath,psfrag,subfigure,scrpage2,slashed,bbold}
\allowdisplaybreaks
\clearscrheadfoot
\ihead{\headmark}

\def\OMIT#1{}

\newcommand{\nn}{\nonumber}

\newcommand{\bea}{\begin{eqnarray}}
\newcommand{\eea}{\end{eqnarray}}

\newcommand{\gsim}{\mathrel{\rlap{\lower4pt\hbox{\hskip1pt$\sim$}}\raise1pt\hbox{$>$}}}

\newcommand{\be}{\begin{equation}}
\newcommand{\ee}{\end{equation}}

\def\lsim{\mathrel{\rlap{\lower4pt\hbox{\hskip1pt$\sim$}}
    \raise1pt\hbox{$<$}}}                % less than or approx. symbol
\def\gsim{\mathrel{\rlap{\lower4pt\hbox{\hskip1pt$\sim$}}
    \raise1pt\hbox{$>$}}}                % greater than or approx. symbol

\begin{document}

\preprint{ACFI-T14-02}

\setlength\baselineskip{17pt}

%%%%%%%%%%%%%%%%%%%%%%%%%%%%%%%%%%%%%%%%%%
%Define Title, Author, Address, Preprint#

\title{\bf  Distinguishing axions from generic light scalars \\using EDM and fifth-force experiments}

\author{ Sonny Mantry}
%\email{mantry147@gmail.com}
\affiliation{High Energy Division, 
                  Argonne National Laboratory, 
                  Argonne, IL 60439}
\affiliation{Department of Physics and Astronomy, 
                   Northwestern University,
                   Evanston, IL 60208}

\author{Mario Pitschmann}
%\email{jqiu@bnl.gov}
\affiliation{Institute of Atomic and Subatomic Physics, 
Vienna University of Technology, 
Stadionallee 2, A-1020 Vienna}                 
                  
\author{Michael J. Ramsey-Musolf}
\email{mjrm@physics.umass.edu}
\affiliation{Amherst Center for Fundamental Interactions, 
		Department of Physics, 
                   University of Massachusetts Amherst, Amherst, MA 01003}
\affiliation{California Institute of Technology, 
		Pasadena, CA 91125}

%\date{\today\\ \vspace{1cm} }

%%%%%%%%%%%%%%%%%%%%%%%%%%%%%%%%%%%%%%%%%%
%Create the title page

\newpage
%\makebox[6.5in][r]{\hfill ANL-HEP-PR-13-18}

%\vskip1.40cm

\begin{abstract}
  \vspace*{0.3cm}

We derive electric dipole moment (EDM) constraints on possible new macroscopic time reversal- and parity violating (TVPV) spin-dependent forces. These constraints are compared to those derived from direct searches in fifth-force experiments and from combining laboratory searches with astrophysical bounds on stellar energy loss. For axion-mediated TVPV spin-dependent forces, EDM constraints dominate over fifth-force limits by several orders of magnitude. However, we show that for a generic light scalar, unrelated to the strong CP problem, present bounds from direct fifth-force searches are  more stringent than those inferred from EDM limits, for the interaction ranges explored by fifth-force experiments. Thus, correlating observations in EDM and fifth-force experiments could help distinguish axions from more generic light scalar scenarios.
%the correlation between EDM and fifth-force limits is significantly different. Depending on the region of parameter space, EDM and  fifth-force limits can be competitive or one can dominate over the other. In fact for forces mediated by a generic scalar and in typical interaction ranges explored by fifth-force experiments, fifth-force limits are more stringent  than EDM limits.
\end{abstract}

\maketitle

\newpage
%\tableofcontents

\section{Introduction}  
  
Tests of the fundamental discrete symmetries of charge conjugation (C), parity (P), and time-reversal (T) have played a vital role in developing the underlying structure of the Standard Model (SM). For example, the discovery of parity-violation led to the formulation of the electroweak sector of the SM as a chiral gauge theory. 
%where fermonic matter fields are understood as appropriate combinations of left- and right-handed chiral multiplets in different representations of the electroweak gauge group. 
The phenomena of CP violation or equivalently T violation, as dictated by the CPT theorem for local quantum field theories, has been extensively studied in various systems within the SM and beyond, and studies of CP-violating observables in the kaon and $B$-meson systems are consistent with expectations based on the CP-phase in the Cabibbo-Kobayashi-Maskawa (CKM) matrix. Nevertheless, 
the observation of  T- and P-violating (TVPV) effects may be indicative of new interactions arising from microscopic P- and CP-violating dynamics going beyond those associated with the CKM CP-violation. 

One interesting scenario is the possibility of a macroscopic spin-dependent (SD) force arising from a light mediator particle associated with physics beyond the SM, where macroscopic is understood as corresponding to an interaction range $r\gg 1$ \AA. As emphasized in the seminal paper by Moody and Wilczek \cite{Moody:1984ba}, a natural candidate for this mediator is the axion, though in the more general case it need not be. 
Through its CP-odd couplings, the same mediator particle can also induce non-zero electric dipole moments (EDMs) of electrons, nucleons, atoms and molecules (for a recent review, see Ref.~\cite{Engel:2013lsa}). It is, then, interesting to ask to what extent dedicated searches for a macroscopic, TVPV SD \lq\lq fifth-force" and EDMs provide complementary probes of this scenario. In this paper, we attempt to address this question.  
%In this paper, we exploit this connection and explore EDM tests as a complementary probe of new macroscopic SD forces. 

A host of \lq\lq fifth-force'" experiments devoted to direct searches of new TVPV SD forces have reported null results \cite{Adelberger:2006dh,Geraci:2008hb,Bordag:2001qi,Decca:2005qz,Nesvizhevsky:2007by,Abele:2003ga,Lamoreaux:1996wh,Klimchitskaya:2001zz,Horvat:2011jh}, while ongoing work seeks to increase the level of sensitivity. For example, one of the more recent techniques \cite{Nesvizhevsky:2005ss} looks for a shift in the spectrum of gravitational quantum states of ultracold \lq\lq bouncing" polarized neutrons that can arise from new SD forces. In another set of experiments, a search for NMR frequency shifts is performed when an unpolarized mass is moved near and far from an ensemble of polarized $^{129}$Xe and $^{131}$Xe gas \cite{Bulatowicz:2013hf}, or polarized $^3$He gas \cite{Tullney:2013wqa,Chu:2012cf}. An overview of various fifth-force experiments can be found in Ref.~\cite{Antoniadis:2011zza}.

In this work, we consider the possibility that a TVPV SD force is mediated by a neutral light spin-zero particle $\varphi$ that interacts with quarks of flavor $q=u,d$ through the 
Lagrangian
\bea
\label{axionquark}
{\cal L}_{\varphi qq} &=& \varphi \>\bar{q} \big(g_s^q+ig_p^q\gamma^5\big) q\ \ \ .
\eea  
These quark-level couplings, in turn, induce the effective scalar and pseudoscalar couplings  to the nucleons ($N$) denoted by $g_s$ and $g_p$ respectively
\bea
\label{axionnucleon}
{\cal L}_{\varphi NN} &=& \varphi \>\bar{N}\big( g_s + ig_p\gamma^5 \big) N\ \ \ . 
\eea
For simplicity, we assume the aforementioned interactions are purely isoscalar $g_{s,p}^u=g_{s,p}^d$. 
The resulting, non-relativistic nucleon-nucleon \lq\lq monopole-dipole" potential is\cite{Moody:1984ba}
\bea
\label{potential}
V(r) &=& g_s g_p \>\frac{\vec{\sigma}_2\cdot \hat{r}}{8\pi M_2} \Big ( \frac{m_\varphi}{r} + \frac{1}{r^2}\Big )e^{-m_\varphi r}\ \ \ ,
\eea
where $\vec{\sigma}_2$ acts on the spin of the polarized nucleon and  $\hat r = \vec r/r$ is the unit vector from the unpolarized object to the polarized particle. Direct searches in fifth-force experiments constrain the strength and range of this potential, giving rise to upper limits on the product of couplings $g_s g_p$ as a function of $m_\varphi$. A summary and detailed discussion of such limits from various experiments using different techniques can be found in Ref. \cite{Antoniadis:2011zza}. 

Since the interaction in Eq.~(\ref{potential}) is TVPV, it will also induce permanent  EDMs of nucleons, nuclei, and diamagnetic atoms\footnote{One could also extend the discussion to consider the interaction of $\varphi$ with leptons and the corresponding implications for paramagnetic systems. Here, we restrict our attention to purely hadronic interactions.}. A non-zero  EDM for an elementary fermion $\psi$ arises from a term in the Lagrangian of the form
\bea
{\cal L} &=& - i\> \frac{d}{2}\> \bar{\psi} \sigma^{\mu \nu} \gamma^5 \psi \>F_{\mu\nu}\ \ \ .
\eea
In the non-relativistic limit, it gives rise to the Hamiltonian
\bea
H &=& -d \> \vec{E}\cdot \frac{\vec{S}}{S}
\eea
for a particle of spin $\vec{S}$ in an electric field $\vec{E}$.  For a non-zero value of $d$,  CP violation is apparent from the CPT theorem and the behavior of the Hamiltonian under time-reversal $T(\vec{E}\cdot \vec{S})=-\vec{E}\cdot \vec{S}$. The current $90\%$ C.L. bounds for the EDM of the neutron,  electron\footnote{The bound on $d_e$ is obtained from the EDM of the ThO molecule assuming the electron EDM would be the only source of any effect. In general, the ThO EDM, as well as that of other paramagnetic systems, may receive significant contributions from a scalar quark $\times$ pseudoscalar electron interaction. A model independent analysis of the most sensitive paramagnetic atomic and molecular EDM experiments then leads to roughly a factor of ten weaker bound on $d_e$.} , and (diamagnetic) Mercury atom are
 \bea
 \label{EDMlimits}
 |d_n| &<& 2.9 \times 10^{-13} \>\text{e fm  \cite{Baker:2006ts}}\ \ \ , \nn \\
 |d_e| &<& 8.7 \times 10^{-16}\>\text{e fm \cite{Baron:2013eja}}\ \ \ , \nn \\
 |d_{Hg}| &<& 2.6 \times 10^{-16}\>\text{e fm \cite{Griffith:2009zz}}\ \ \ ,
 \eea
 (for a review, see {\em e.g.} Ref. \cite{Engel:2013lsa}).

Null results for EDM searches generally imply severe constraints on TVPV interactions, so it is interesting to investigate the implications of EDM searches for the interpretation of fifth-force designed to probe the interaction inEq.~(\ref{potential}). It is well-known that when $\varphi$ is the axion($a$) \cite{Peccei:1977hh,Peccei:1977ur,Weinberg:1977ma,Wilczek:1977pj}, invoked to solve the strong CP problem, EDM constraints on $g_s g_p$ are several orders of magnitude more stringent~\cite{Rosenberg:2000wb} than those derived from fifth-force experiments. As we discuss below, this situation results from the unique properties of the axion as a pseudo Goldstone boson of spontaneously broken Peccei-Quinn symmetry. On the other hand, when $\varphi$ is a generic spin-zero particle, the relative impact of EDM and fifth force searches depend strongly on $m_\varphi$. Thus, the two classes of experiments  provide complementary probes. Should either type of search (or both) yield a non-zero result, then one could infer information about the  existence and nature of the $\varphi$, its couplings to matter, and its mass.  The key relevant differences between axions and generic scalars are summarized in Table \ref{axion-generic} and the details are explained in the subsequent sections.
\begin{table}
\renewcommand{\arraystretch}{1.5}
\begin{center}
\begin{tabular}{|c|c|c|}
\hline
 Properties & Axion ($a$)  &  Generic Scalar   ($\varphi$)  \\
\hline\hline
$\qquad${\rm  Leading source of EDM}$\qquad$&$\qquad$TVPV quark mass term $\qquad$&$\qquad$dynamical $\varphi$  exchange $\qquad$ \\[-1ex]
&$\sim \bar{\theta} \>m_q\> \bar{q} \>i\gamma_5  \> q$ & \\[1ex]
\hline
$g_s $& $\sim \bar{\theta} \>\frac{m_q}{f_a}\propto \bar{\theta} \>m_a$  & arbitrary/unrelated to $m_\varphi$  \\[1ex]
\hline
$g_p $& $\sim \>\frac{m_q}{f_a}\propto \>m_a$  & arbitrary/unrelated to $m_\varphi$  \\[1ex]
\hline
$g_s g_p$& $\sim \bar{\theta} \>\frac{m_q^2}{f_a^2}\propto \bar{\theta} \>m_a^2$  & arbitrary/unrelated to $m_\varphi$ \\[1ex]
\hline
\end{tabular}
\end{center}
\caption{For the case of the axion, a non-zero EDM arises from TVPV quark mass terms that are induced when eliminating the QCD $\bar{\theta}$-term  via an axial $U(1)_A$ rotation. The current EDM bounds on these TVPV quark mass terms imply $|\bar{\theta}| < 10^{-10}$.  The product of couplings $g_s g_p \sim  \bar{\theta} \>\frac{m_q^2}{f_a^2}$ are  proportional to the same $\bar{\theta}$-parameter and $f_a$ denotes the Peccei-Quinn symmetry breaking scale and is related to the axion mass as $m_a \propto 1/f_a$. Thus, the EDM bound $|\bar{\theta}| < 10^{-10}$ implies severe constraints on $g_s g_p$ which dominate over fifth-force constraints. By contrast for a generic scalar, unrelated to a solution to the Strong CP problem, the EDM is generated by dynamical $\varphi$-exchange between quarks or nucleons and the product $g_s g_p$ is unrelated to the $\bar{\theta}$-parameter. Thus, EDM constraints have a much weaker impact on $g_sg_p$ for a generic scalar and fifth-force limits dominate for the range of interactions they probe. }
\label{axion-generic}
\end{table}

In what follows, we provide a rationale for these observations. In Sections \ref{EDMSM} and \ref{axionphysics}, we review  strong CP-violation in the Standard Model and the axion mechanism invoked to solve the Strong CP-problem. Although this discussion is not new, a brief pedagogical discussion is useful as a means of setting the stage for the generic $\varphi$ scenario and of elucidating the distinct EDM constraints on the interaction Eq.~(\ref{potential}) for the axion and generic $\varphi$ cases. In Section \ref{generic} we consider the generic $\varphi$ scenario in detail. In particular, we derive order of magnitude bounds on $g_s g_p$ from limits on the $^{199}$Hg EDM and show that for $m_\varphi<<m_\pi$, the fifth force constraints are several orders of magnitude stronger. Our approach in this instance is to obtain benchmark, order of magnitude estimates for the EDM constraints rather than to carry out an exhaustive computation of all possible $\varphi$ contributions that would require extensive nuclear many-body computations. Nevertheless, we endeavor to be as complete as possible wherever analytic computations are tractable. The corresponding technical details appear in a set of four Appendices that follow our conclusions in Section \ref{sec:conclude}.

\section{Strong CP-Violation in the Standard Model}
\label{EDMSM}

Within the SM,  two sources of CP violation can generate a non-zero EDM.  The first arises from the complex phase in the CKM matrix that characterizes the strength of flavor changing charged currents. CP violation associated with the CKM matrix has been confirmed and studied in great detail through the mixing and decay properties of K- and B-mesons. The contribution of the CKM phase to the neutron EDM is of order $d_n \sim 10^{-32}$ $e$ cm \cite{Nanopoulos:1979my,Nanopoulos:1979xt,Ellis:1976fn,Shabalin:1979gh,Khriplovich:1981ca,Deshpande:1981mp,Czarnecki:1997bu,Eeg:1983mt,Dar:2000tn}, about six orders of magnitude below the current experimental limit. As a result, CKM induced effects give a negligible background to present and prospective EDM searches.

%%%OMIT
\OMIT{The dominant contribution comes from the $\Delta S=1$ hadronic weak interaction and shows up with long range contributions in terms of a pseudoscalar loop as well as a short range counterterm contribution at the level of the hadronic effective field theory.}
%%%

The second source of CP violation in the SM arises from the CP violating term in the QCD Lagrangian
\bea
\label{thetaterm}
{\cal L}_{QCD}^{CPV} &=& \bar{\theta}\> \frac{\alpha_s}{16 \pi}\> G_{\mu \nu}^a \tilde{G}^{a\mu \nu}\ \ \ ,
\eea
where $\tilde{G}_{\mu \nu}^a=\varepsilon_{\mu\nu\rho\sigma} G^{a,\rho \sigma}$. The parameter $\bar{\theta}$ is given by
\bea
\label{thetabar}
\bar{\theta} = \theta + \text{arg}(\text{det} \>M_q')\ \ \ ,
\eea
where the $\theta$-parameter arises from the non-trivial structure of the QCD vacuum and $M_q'$ corresponds to the original non-diagonal quark mass matrix after electroweak symmetry breaking. Such a term is not forbidden by any symmetry and is in fact expected due to the non-trivial structure of the QCD vacuum, the anomaly in the axial $U(1)_A$ transformation on quarks \cite{Adler:1969gk, Bell:1969ts}, and the absence of any massless quark in the SM. This term corresponds to a source of {flavor-diagonal} CP violation, as opposed to the CKM phase associated with flavor-changing CP violation.

The existence of gauge-equivalent vacuum instanton configurations, with distinct topological properties, requires the QCD vacuum to be given by a gauge-invariant superposition of these configurations. Each such vacuum state is labeled by a $\theta$-parameter
\bea
|\theta\rangle = \sum_{n} e^{i n\theta} |n\rangle\ \ \ ,
\eea
where $n$ denotes the topological winding number of the instanton configuration corresponding to the vacuum state $|n\rangle$. This non-trivial structure of the QCD vacuum is accounted for by the $\theta$-term in Eqs.~(\ref{thetabar}) and (\ref{thetaterm}).
There exists a connection between the QCD $\theta$-vacuum and the axial $U(1)_A$ anomaly. The axial $U(1)_A$ transformation corresponds to a  phase rotation of a quark field given by
\bea
\label{U1A}
\psi \to e^{-i\alpha \gamma^5}\>\psi\,, \qquad \bar{\psi} \to \bar{\psi}\>e^{-i\alpha \gamma^5}\ \ \ ,
\eea
where $\alpha$ denotes the  phase rotation angle. This transformation is a classical symmetry of the Lagrangian in the limit of massless quarks. However, it is anomalous at the quantum level. The divergence of the current 
\bea
j_\mu^5 &=& \bar{\psi} \gamma_\mu \gamma^5\psi\ \ \ ,
\eea
associated with the  $U(1)_A$ transformation,  is given by
\bea
\label{U1Adiv}
\partial^\mu j^{5}_{\mu} &=& 2i  m_q\>\bar{\psi} \gamma^5 \psi + \frac{\alpha_s}{8\pi} \>G_{\mu \nu}^a \tilde{G}^{a\mu \nu}\ \ \ .
\eea
We see that the quark masses explicitly break the $U(1)_A$ symmetry of the Lagrangian even at the classical level. The second term in Eq.~(\ref{U1Adiv}), with the same structure as the QCD CP violating term in Eq.~(\ref{thetaterm}), is the result of the anomaly and arises from the non-trivial Jacobian  in the QCD path-integral \cite{Fujikawa:1979ay, Fujikawa:1980eg, Fujikawa:1980vr, Fujikawa:1980rc} that arises from the transformation in Eq.~(\ref{U1A}) 
\bea
\label{jacobian}
{\cal D}\psi{\cal D}\bar{\psi}  \to {\cal D}\psi {\cal D}\bar{\psi}\> \text{Exp} \Big [ 2i\alpha \int d^4x\>  \frac{\alpha_s}{16 \pi}\> G_{\mu \nu}^a \tilde{G}^{a\mu \nu}\Big ]\ \ \ .
\eea 
For a $U(1)_A$ transformation of a massless quark, the only effect of the axial $U(1)_A$ transformation in Eq.~(\ref{U1A}) is to shift the value of the $\theta$-parameter 
\bea
\label{thetashift}
\theta \to \theta + 2\alpha\ \ \ .
\eea
Since the $U(1)_A$ transformation just amounts to a change of variables on the QCD path integral, the shift in Eq.~(\ref{thetashift}) implies that the path integral cannot depend on $\theta$, rendering it an unphysical parameter. Thus, if there is at least one massless quark, the QCD CP violating term can be completely removed by an appropriate $U(1)_A$ phase rotation.

However, it is now well-established that there are no massless quarks in the SM \cite{Leutwyler:2009jg}. In addition to the shift in the $\theta$-parameter, the $U(1)_A$ transformation then also changes the phase of the quark mass. In this case, the $U(1)_A$ transformation cannot be used to eliminate the CP violating effect in QCD. Instead, it can only move the effect between the $G\widetilde{G}$ and the quark mass operators.

In fact, a flavor-diagonal $U(1)_A$ transformation can be used to remove the overall phase in the quark mass matrix so that all of the flavor-diagonal CP violation is contained in the $\bar{\theta}$-term in Eq.~(\ref{thetaterm}), where $\bar{\theta}$ is given by Eq.~(\ref{thetabar}). Alternatively, one can perform an axial $U(1)_A$ rotation to eliminate the $\bar{\theta}$-term so that the flavor-diagonal CP violation effect is contained entirely in CP violating quark mass terms. Integrating out the heavy quarks $c$, $b$ and $t$ one has
\bea
\label{massCPV}
{\cal L}_{CPV} &=& i\bar{\theta}\> \frac{m_u m_d m_s}{m_um_d+m_um_s+m_dm_s}\> \big(\bar{u} \gamma^5 u+\bar{d} \gamma^5 d+\bar{s} \gamma^5 s \big) \ \ \ .
\eea
Note that this term is proportional to the product of quark masses so that in the presence of a massless quark, there is no flavor-diagonal CP violation as expected.

Given that the contribution of the CKM phase to EDMs in the SM are negligibly small, the observation of a non-zero EDM can be interpreted as arising from CP-violating mass term in Eq.~(\ref{massCPV}) or equivalently from the $\bar{\theta}$-term in Eq.~(\ref{thetaterm}). The current limits on $d_n$ and $d_{Hg}$, translate into the bound
\bea
|\>\bar{\theta}\>| \lesssim 10^{-10}\ \ \ . 
\eea
The absence of a SM  explanation for such a small value of $\bar{\theta}$ corresponds to the  well-known Strong CP problem. 

 %%%%%%%%%%%%%%%%%%%%%%%%%%%%%%%%%%%%
\section{EDMs and Spin-Dependent Forces via Axions} 
\label{axionphysics}

The generation of non-zero EDMs in the SM, through either the CKM phase or the $\bar{\theta}$-term (or both), is not in general associated with a macroscopic SD force.  Such an association can, however, arise in scenarios beyond the SM that involve a light mediator particle with CP-violating couplings to SM fermions. A well-known example of such a light mediator particle is the axion, introduced to provide a dynamical explanation of the strong CP problem.
Here we give a brief overview of the axion mechanism that can then be contrasted with the case of a more general scalar mediator considered in this work. In particular, we will show that the relative implications of EDM and fifth-force constraints are quite distinct for the axion and generic scalar cases. More comprehensive and detailed reviews on axion physics can be found, for example, in Refs. \cite{Pospelov:2005pr,Kim:2008hd}.

For the purposes of illustration, we consider the axion mechanism in the Kim-Shifman-Vainstein-Zakharov (KSVZ) model \cite{Kim:1979if,Shifman:1979if}. In this model, the SM is augmented by a new massless electroweak-singlet quark $\psi$ and a complex scalar $\Phi$
\bea
\label{AxionLagr}
\delta {\cal L} &=& \partial_\mu \Phi^\dagger \partial^\mu \Phi + \mu_\Phi^2 \Phi^\dagger \Phi - \lambda_{\Phi}  (\Phi^\dagger \Phi)^2 + \bar\psi i\slashed\partial\psi + y\>\bar\psi_R\Phi\psi_L + h.c.\ \ \ ,
\eea
where  $\psi_L=\frac{1}{2}(1 - \gamma^5)\psi$ and $\psi_R=\frac{1}{2}(1 + \gamma^5)\psi$ denote the left-handed and right-handed chiral components of the new massless quark respectively. The Lagrangian $\delta {\cal L}$ is invariant under a global chiral $U(1)_{\text{PQ}}$ Peccei-Quinn transformation 
\bea
\label{PQ}
\psi \to e^{-i\alpha \gamma^5}\>\psi\,, \qquad \bar\psi\to\bar\psi\>e^{-i\alpha \gamma^5}\,, \qquad  \Phi \to e^{-2i\alpha}\>\Phi\ \ \ .
\eea
The SM fields are neutral under  $U(1)_{\text{PQ}}$,  so that the full theory Lagrangian is invariant under this transformation at the classical level.
However, as in the case of the axial $U(1)_A$ transformation, the Peccei-Quinn transformation is anomalous and contributes a shift to the value of $\theta$,  as shown in Eqs.~(\ref{jacobian}) and (\ref{thetashift}). Thus, by an appropriate $U(1)_{\text{PQ}}$, one can completely rotate away the $\bar{\theta}$-parameter, thereby solving the Strong CP problem. 

Since a massless, electroweak-singlet quark is not observed in nature, the $U(1)_{\text{PQ}}$ symmetry of the Lagrangian must be spontaneously broken at a high enough scale $f_a$ so that the new quark acquires a large enough mass to avoid current experimental limits. The spontaneous symmetry breaking occurs via the vacuum expectation value
\bea
\langle \Phi \rangle = f_a\ \ \ ,
\eea
and the excitations about this ground state value can be written as
\bea
\label{Phi}
\Phi(x) &=& \frac{f_a + \rho(x)}{\sqrt{2}}\> e^{i \mathrm a(x)/f_a}\ \ \ .
\eea
The heavy field $\rho(x)$ corresponds to radial excitations and $\mathrm a(x)$ is the axion corresponding to the Goldstone boson associated with the spontaneous symmetry breaking of $U(1)_{\text{PQ}}$. However, since the $U(1)_{\text{PQ}}$ symmetry is explicitly broken by the chiral anomaly, the axion is a \textit{pseudo}-Goldstone boson and acquires a potential and a non-zero mass.
Experimental constraints imply that  $10^9 \lsim f_a \lsim 10^{12}$ GeV, which constitute the \lq\lq axion-window" \cite{Kim:2008hd}. 

After the spontaneous symmetry breaking, the new electroweak-singlet quark acquires a large mass $m_\psi \sim f_a$ via its Yukawa interaction with $\Phi$ in Eq.~(\ref{AxionLagr}). The field $\rho(x)$ in Eq.~(\ref{Phi}) also acquires a large mass. One can construct a low energy effective theory by integrating out the heavy fields $\psi(x), \rho(x)$, where the low energy degrees of freedom correspond to SM fields and the axion. The general form of such an effective theory is obtained by observing the symmetry properties of the full theory. Note that the $U(1)_{PQ}$ transformation in Eq.~(\ref{PQ}) results in the shifts
\bea
\bar{\theta} \to \bar{\theta} + 2\alpha\,, \qquad \frac{\mathrm  a(x)}{f_a} \to \frac{\mathrm a(x)}{f_a} -2\alpha\ \ \ ,
\eea
so that the quantity $\bar{\theta} + \frac{\mathrm a(x)}{f_a} $ is left invariant. This implies that all axion interactions in the effective theory must be formulated in terms of this invariant combination as a fundamental building block. In particular, the $\bar{\theta}$-parameter in Eq.~(\ref{thetaterm}) must be replaced as
\bea
\label{PQinv}
 \bar{\theta} \to \bar{\theta} + \frac{\mathrm  a(x)}{f_a}\ \ \ ,
\eea
so that the $\bar{\theta}$ parameter is effectively promoted to a dynamical field.
The effective interaction Lagrangian for the axion now takes the general form 
 \bea
 \label{La1}
 {\cal L}_{a} &=& \frac{\alpha_s}{16 \pi}\> \Big (\bar{\theta} + \frac{\mathrm  a}{f_a} \Big)\>   G_{\mu \nu}^a \tilde{G}^{a\mu \nu}  - m_q\> \bar{q}q +\cdots\ \ \ .
 \eea
 where the \lq\lq $+\cdots$" denote the axion kinetic and mass terms as well as possible higher-dimension axion interactions. 
%Note that in general additional terms from higher dimension axion interactions  are also possible and the kinetic term for the axion $\frac{1}{2}\>\partial^\mu \mathrm  a(x) \partial_\mu \mathrm  a(x)$ is implicit. 
Note that we have included the quark mass term in the definition ${\cal L}_a$ since, as discussed below, an axial $U(1)_A$ transformation can move the axion coupling entirely into  the quark mass term. For purposes of illustration, we work in QCD with one quark flavor.

The couplings of the axion to the SM matter fermions can be made more explicit by rotating $\bar{\theta}$-term in Eq.~(\ref{thetaterm}) into the quark mass matrix before introducing axions by the replacement  in Eq.~(\ref{PQinv}). Prior to introducing the axion, the relevant terms in the Lagrangian of  QCD with a single quark flavor are
\bea
{\cal L} &=& \bar\theta\>\frac{\alpha_s}{16\pi}\> G^a_{\mu \nu}\tilde{G}^{a\mu\nu}  - m_q\>\bar qq\ \ \ .
\eea
Performing an axial $U(1)_A$ transformation to rotate the $\bar{\theta}$-term into the quark mass, the Lagrangian can be brought into the form 
\bea
\label{massCPV1}
{\cal L} &=& - m_q\cos\bar\theta\>\bar qq + m_q\sin\bar\theta\>\bar qi\gamma^5q\ \ \ ,
\eea
which will reproduce the analog of the term in Eq.~(\ref{massCPV}), when expanded to leading power in $\bar{\theta}$ and generalized to three quark flavors \cite{Baluni:1978rf}. Inclusion of the axion interactions in the effective theory can now be obtained by implementing the replacement in Eq.~(\ref{PQinv}), leading to
\bea
\label{La2}
{\cal L}_a &=& - \cos\Big(\bar\theta +\frac{\mathrm  a}{f_a}\Big)\>m_q\>\bar qq + m_q\sin\Big(\bar\theta +\frac{\mathrm a}{f_a}\Big)\>\bar qi\gamma^5q\ \ \ ,
\eea
which is equivalent to the form in Eq.~(\ref{La1}). The form of Eq.~(\ref{La2}) makes manifest the couplings of the axion to the SM quark. In general the axion can acquire a non-zero vacuum expectation value (vev) so that
\bea
\mathrm a(x) &=& \langle \mathrm a\rangle + a(x)\ \ \ ,
\eea
where $a(x)$ denotes the axion field corresponding to excitations above the vev  $\langle \mathrm a\rangle$. After the axion acquires a non-zero expectation value, the new \textit{induced} $\bar{\theta}$ parameter ($\theta_{\text{eff}}$) is given by
\bea
\theta_{\text{eff}} &=& \bar{\theta} + \frac{\langle \mathrm a\rangle}{f_a}\ \ \ ,
\eea
so that the axion Lagrangian in Eq.~(\ref{La2}) can be brought into the form
\bea
\label{La3}
{\cal L}_a &=& - \cos \Big (\theta_{\text{eff}} +\frac{a}{f_a} \Big )\>m_q\>\bar qq + m_q\sin\Big(\theta_{\text{eff}} +\frac{ a}{f_a}\Big)\>\bar qi\gamma^5q\ \ \ .
\eea
An axion potential is generated through non-perturbative QCD effects which generate a quark condensate so that
\bea
V\Big(\theta_{\text{eff}} + \frac{a}{f_a}\Big) &=& \>-\chi(0)  \cos \Big (\theta_{\text{eff}} +\frac{a}{f_a} \Big)\ \ \ ,
\eea
where the {topological susceptibility}  is given by
\bea
\chi(0) = -m_q\>  \langle \bar qq \rangle\ \ \ .
\eea
Generally, the ground state axion potential, when expanded around its minimum, has the form
\bea
\label{axionpot}
V(\theta_{\text{eff}}) \>\simeq \>\frac{1}{2}\> \chi (0) \> \theta_{\text{eff}}^2\ \ \ .
\eea
Since the minimization of the ground state axion potential requires $\theta_{\text{eff}}=0$, there is no flavor-diagonal CP violation and a correspondingly vanishing contribution to the EDM. In this way, dynamical relaxation in the ground state axion potential solves the strong CP problem and eliminates flavor-diagonal CP-violation. 

The presence of additional higher-dimensional CP-odd operators, such as the quark chromo-electric dipole moment,  can generate terms that are linear in $\theta_{\text{eff}}$ in the axion potential. This can occur via mixed correlators of the form \cite{Pospelov:2005pr}
\bea
\chi_{\text{CP}} (0) &=& -i\> \text{lim}_{k\to 0}\> \int d^4x \> e^{ik\cdot x} \langle 0 | T(G\tilde{G} (x), {\cal O}_{\text{CP}} (0)) | 0 \rangle\ \ \ .
\eea
%An example of such a CP-odd operator is the quark chromoelectric dipole moment ${\cal O}_{\text{CP}}=\tilde{d}_q \bar{q} G^{\mu \nu}\sigma_{\mu \nu}\gamma^5 q$. 
Such mixed correlators can give rise to an axion potential of the form
\bea
\label{axionpot1}
V(\theta_{\text{eff}}) &\simeq& \chi_{\text{CP}}(0) \>\theta_{\text{eff}}  + \frac{\chi(0)}{2}\>\theta_{\text{eff}}^2\ \ \ .
\eea
In this case, the potential is minimized at non-zero value of $\theta_{\text{eff}}$ given by 
\bea\label{thetadef}
\theta_{\text{eff}}  = -\frac{\chi_{\text{CP}}(0)}{ \chi (0)}\ \ \ ,
\eea
resulting in a non-vanishing contribution to EDMs\footnote{This non-vanishing $\theta_{\text{eff}}$ corresponds to $\theta_{\text{ind.}}$ in the notation of Ref. \cite{Pospelov:2005pr}.}.

Expanding the Lagrangian in Eq.~(\ref{La3}) in $\theta_{\text{eff}}$  and $a(x)$, gives the result
\bea\label{La4}
{\cal L}_a &=&  \Big ( \frac{\theta_{\text{eff}} }{f_a} a-1 \Big )\>m_q\>\bar qq + \Big (\theta_{\text{eff}} + \frac{a}{f_a}\Big)\>m_q\> \bar qi\gamma^5q + \frac{m_q}{2f_a^2}\>a^2\>\bar qq 
 + \cdots\ \ \ .
\eea
This form of the Lagrangian makes explicit the scalar ($g_{a,s}^q$) and pseudoscalar  ($g_{a,p}^q$)  couplings and the induced mass $m_a$ of the axion
\bea
\label{gasp}
g_{a,s}^q &=& \frac{\theta_{\text{eff}} m_q }{f_a}\,, \qquad g_{a,p}^q = \frac{m_q }{f_a}\,, \qquad m_a \simeq \frac{1}{f_a} |\chi (0)|^{1/2}\ \ \ .
\eea
Note that the CP-odd mass term $\theta_{\text{eff}} m_q\> \bar qi\gamma^5q $ in Eq.~(\ref{La4}) is the analogue of Eq.~(\ref{massCPV}) for the case of one quark flavor. 
Moreover, since $f_a\gg |\chi (0)|^{1/4}$, the axion is very light and can mediate a macroscopic SD force. Based on the axion couplings to the quark, the product of couplings in the corresponding potential in Eq.~(\ref{potential}) is expected to be proportional to the product of the scalar and pseudoscalar axion couplings to the quark
\bea
\label{gs1gp2a}
g_s^q g_p^q \propto \theta_{\text{eff}}\> \frac{m_q^2}{f_a^2}\ \ \ ,
\eea
with the constant of proportionality being determined by the nuclear/nucleon matrix elements relevant to the test objects in the experiment. Note that the size of the SD fifth-force induced by the axion is heavily suppressed by the factor of $m_q^2 /f_a^2$.

The dominant contribution of the axion to EDMs will come from a matrix element involving the CP-odd quark mass term $m_q\theta_{\text{eff}} \bar qi\gamma^5q$ in Eq.~(\ref{La4}). Note that in this case, the suppression factor $m_q^2 /f_a^2$, present in the macroscopic SD fifth-force, is absent. As a result, EDM constraints on $\theta_{\text{eff}}$ dominate over the constraints from fifth-force experiments by several orders of magnitude. 

EDM bounds require $\theta_{\text{eff}} \lsim 10^{-10}$, so that for quark masses $m_q \sim 1 \text{ MeV}$ and a   Peccei-Quinn scale $f_a \sim 10^9 - 10^{12} \text{ GeV}$, the coupling $g_{a,s}^q $  must lie below $10^{-25} - 10^{-22}$. Correspondingly, the bound on the pseudoscalar coupling is  $g_{a,p}^q < 10^{-15} - 10^{-12}$.
The resulting product of the macroscopic couplings in Eq.~(\ref{potential}), for the  fifth-force potential due to an axion mediator, are bounded from EDM constraints as
\bea
\label{eq:axionbound}
g_s g_p \propto \theta_{\text{eff}}\> \frac{m_q^2}{f_a^2} < 10^{-40} - 10^{-34}\ \ \ .
\eea
These EDM bounds are the the most stringent constraints; in fact even stronger than those derived by combining the existing fifth-force laboratory limits with astrophysical limits from SN 1987A (see bottom panel in Fig.~4 of \cite{Raffelt:2012sp}). As we discuss below, this situation contrasts sharply with the case of a generic scalar, for which $g_s$ and $g_p$ are {\em a priori} unrestricted free parameters and unrelated to the strong CP parameter $\theta_{\text{eff}}$.

%%%%OMIT
\OMIT{
Furthermore, we would like to mention, while we have considered here the simplest axion model, which is the Kim-Shifman-Vainstein-Zakharov (KSVZ) model, there are several other axion models. For example, in the original Peccei-Quinn-Weinberg-Wilczek (PCWW) model, the axion degree resides in a combination of the phases of the Higgs doublets in the standard model. Hence, the axion appears already in the SM quark fields and no additional heavy quarks are added to the SM quark sector. While the PCWW model has been excluded on experimental grounds, there are still other axion models coming in different versions. For details on those and subtleties concerning the multi-flavor case as well as the inclusion of SUSY we refer to the reviews \cite{?}.
Due to the order of magnitude in the axion-quark couplings, one finds that the sensitivity of current fifth-force experiments falls short \cite{?} by several orders of magnitude  to detect a macroscopic fifth-force due to the axion. In this case, the EDMs themselves provide the strongest bounds on possible new SD macroscopic forces.} 
%%%%END-OMIT

\section{Spin-Dependent Forces and EDMs from a Generic Light Scalar}
\label{generic}

We now turn to the generic light scalar case and return to the basic interactions of Eqs.~(\ref{axionquark},\ref{axionnucleon}). Our objective is to estimate the diamagnetic atom and nucleon EDMs induced by these interactions as functions of the parameters $g_s$ and $g_p$ and derive order-of-magnitude bounds on their product. Before doing so, we comment on the possible origin of the interaction in Eq.~(\ref{axionquark}). Although this interaction does not respect the SM electroweak symmetry, it may be the low-energy remnant of a more complete theory that does so at high scales. Consider, for example, an extension of the SM scalar sector that includes an additional complex gauge singlet. After electroweak symmetry-breaking, the SM Higgs scalar will in general mix with one component of the singlet, unless one imposes a discrete $\bm{Z}_2$ symmetry on the scalar potential. If the electroweak-singlet vacuum also spontaneously breaks CP, then mixed scalar-pseudoscalar states will occur. The SM Yukawa interactions will then give rise to both types of terms in Eq.~(\ref{axionquark}), with $g_{s,p}$ being functions of the quark Yukawa couplings and parameters in the scalar potential. The question, then, is to determine the extent to which EDMs and fifth force experiments might constrain such a scenario if one of the scalars is ultra-light\footnote{In this case, there will in general also exist heavier mixed scalar-pseudoscalar states whose couplings to quarks will also be functions of the Yukawa couplings and scalar potential.} (for a concrete realization, see, {\em e.g.} Ref.~\cite{Barger:2008jx}).

\subsection{EDMs induced by a generic light scalar} 
\label{EDMscalar}

\begin{figure}[h!]
  \centering
  \vspace{1mm}
  \includegraphics[width=0.85\linewidth]{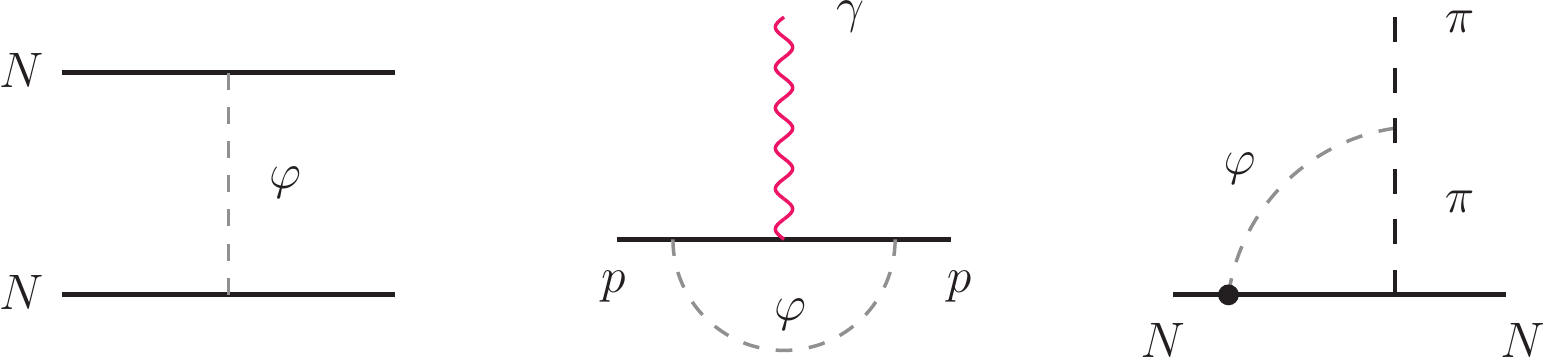}
  \caption{Representative diagrams of the contribution to nuclear EDMs arising from exchanges of the light scalar $\varphi$ that mediates the macroscopic SD force. The first diagram corresponds to $\varphi$ exchanges between nucleons in the nucleus. The second and third diagrams can be interpreted as an induced proton EDM and CP-odd pion-nucleon coupling due to $\varphi$-exchange. 
 %Since $m_\varphi \ll \Lambda_{QCD}$,  the light scalar $\varphi$ cannot be integrated out and must be included as a propagating degree of freedom in nuclear calculations.
  }
  \label{AEDM}
\end{figure}

We identify three classes of effects associated with Eqs.~(\ref{axionquark},\ref{axionnucleon}) that contribute to EDMs, illustrated in Fig.~\ref{AEDM}: (a) direct $\varphi$ exchange between two nucleons that generates the potential (\ref{potential}) and contributes to the nuclear Schiff moments of diamagnetic atoms (first panel) ; (b) $\varphi$ loops involving one factor each of the scalar and pseudo scalar couplings that induce a nucleon EDM (middle panel); and (c) $\varphi$ loops that induce a TVPV $\pi NN$ coupling that, in turn, generates the nuclear Schiff moment {\em via} $\pi$-exchange between two nucleons (third panel). 

%The most sensitive EDM searches have been carried out for the neutron as well as paramagnetic and diamagnetic atoms and molecules. For these systems, the calculation of EDMs in terms of the underlying microscopic sources of CP violation is complicated by the interplay of atomic, nuclear, and hadronic physics. In particular, here we are interested in being able to obtain the dependence of the induced EDMs on the nucleon level couplings  $g_s$ and $ g_p$, given in Eq.~(\ref{axionnucleon}). In particular, we estimate the contribution to the diamagnetic mercury EDM $d_{Hg}$ which is given in terms of the nuclear Schiff moment. The current bound on $ d_{Hg}$, given in Eq.~(\ref{EDMlimits}), will then  translate into bounds on the product of nucleon level couplings $g_sg_p$.

The computation of an EDM of a strongly-interacting many-body system is highly non-trivial, and theoretical subtleties arise at the hadronic, nuclear, and atomic levels (for reviews, see Refs.~\cite{Engel:2013lsa,Pospelov:2005pr,Ginges:2003qt}). Our objective is not to carry out definitive computations of the contributions illustrated in Fig.~\ref{AEDM} that take these subtleties into account, but rather to obtain benchmark estimates that give reasonable indications of the EDM bounds on  $g_sg_p$. To that end, we first observe that the dominant contribution to the nuclear Schiff moment is likely to arise from direct $\varphi$ exchange (first panel of Fig.~\ref{AEDM}). Unfortunately, we do not have at our disposal the machinery needed to carry out a sophisticated many-body computation involving the potential of Eq.~(\ref{potential}). On the other hand, detailed computations of nuclear Schiff moments have been performed assuming a $\pi$-exchange mechanism, where 
one $\pi NN$ vertex is the leading order strong coupling and the other is a TVPV vertex. The leading TVPV $\pi NN$ interaction is given by 
\bea
{\cal L}_{\pi NN} &=& \bar{g}^{(0)}_{\pi NN}\> \bar{N}\tau^a N\pi^a + \bar{g}_{\pi NN}^{(1)}\> \bar{N}N\pi^0 + \bar{g}^{(2)}_{\pi NN}\> \big(\bar{N} \tau^a N \pi^a - 3 \bar{N} \tau^3 N \pi^0\big)\ \ \ ,
\eea
with $ \bar{g}^{(0)}_{\pi NN}, \bar{g}_{\pi NN}^{(1)}$, and $\bar{g}^{(2)}_{\pi NN}$ denote the induced isoscalar, isovector, and isotensor TVPV couplings, respectively. The  nuclear Schiff moment can then be expressed as \cite{Engel:2013lsa}
\begin{align}
\label{schiff}
  S_{Hg}=g_{\pi NN}\>\Big(a_0\>\bar{g}_{\pi NN}^{(0)} + a_1\>\bar g_{\pi NN}^{(1)} +a_2\>\bar g_{\pi NN}^{(2)}\Big)\>e\text{ fm}^3\ \ \ ,
\end{align}
with $g_{\pi NN} = m_n g_A/f_\pi\simeq 13.5$. A compilation of the $a_i$ obtained from various calculations, along with a set of \lq\lq best values" and \lq\lq reasonable ranges" is given in Ref.~\cite{Engel:2013lsa}. For $^{199}$Hg, one has  $0.005 < a_0 < 0.5$, $-0.03 < a_1 < 0.09$, and $ 0.01 < a_2 <0.06$ with the  \lq\lq best" values for the coefficients are $a_0=0.01$, $a_1=\pm 0.02$, $a_2=0.02$. For the $\varphi$ scenario we consider here, only $a_0$ is relevant.
The resulting mercury EDM has the form \cite{deJesus:2005nb,Griffith:2009zz} 
\bea
\label{dHg}
d_{Hg}=d_{Hg}(S_{Hg}[\bar{g}_{\pi NN}^{(i)}])\simeq -2.8\times10^{-4}\>\frac{ S_{Hg}}{\text{ fm}^2}\ \ \ ,
\eea

Under the assumption that the interactions in Eqs.~(\ref{axionquark},\ref{axionnucleon}) are isoscalar, the loop effects associated with the third panel in Fig.~\ref{AEDM} will induce a non-zero value for $ \bar{g}_{\pi NN}^{(0)} $. Apart from the different ranges associated with $\pi$ and $\varphi$ exchange, the effect of $\varphi$ loop-induced TVPV $\pi$-exchange is likely to be suppressed by $1/16\pi^2$ relative to the impact of the direct $\varphi$-exchange potential, implying that the impact of the latter is likely to be two-orders of magnitude stronger than the former. With this context in mind, we are able to obtain tractable estimates of the loop-induced $ \bar{g}_{\pi NN}^{(0)} $ and use them, along with existing computations of the $a_0$ for mercury, to derive a bound on $g_s g_p$. To be conservative, we will then multiply this bound by $10^{-2}$ to take into account the loop suppression relative to direct $\varphi$ exchange and compare the resulting bound with the direct, fifth force limits. As we discuss below, the latter are still several orders of magnitude more stringent than our estimated EDM bound. 

Before proceeding, we comment on the $\varphi$ loop-induced contributions to the nucleon EDMs (Fig.~\ref{AEDM}, middle panel). Since the neutron  is electrically neutral, the leading contribution to $d_n$ involves the magnetic moment insertion. For a consistent calculation, we employ heavy baryon chiral perturbation theory (HBChPT)\cite{Jenkins:1990jv}, which involves expanding about both the chiral ($m_\pi\to 0$) and static nucleon ($m_N\to\infty$) limits. At leading non-trivial order in the heavy baryon expansion (order $q/m_N\sim m_\pi/m_N$), the photon-nucleon coupling is magnetic and, thus, no EDM is generated. At next order, the spin-orbit correction induces a coupling to the electric field, allowing for an EDM to be generated. The resulting $d_n$ contribution is thus second order in $q/m_n$, where $q$ denotes a small momentum or $m_\pi$. A proton EDM can be generated via the electric photon coupling in Fig.~\ref{AEDM} (middle panel). However, the contribution of this diagram to the atomic EDM is suppressed by at least one power of $q/m_N$ relative to that of the diagram in the third panel of Fig.~\ref{AEDM} (see the discussion in appendix~\ref{PEDM}). i.e. we find that the loop-induced  $ \bar{g}_{\pi NN}^{(0)} $ arises at zeroth order in the heavy baryon expansion and gives the dominant contribution at one loop. Consequently, we expect that our strategy for bounding $g_s g_p$ from $d_{Hg}$ as outlined above will yield the most stringent limit. 

The $\varphi$ loop indicated in the third panel of Fig.~\ref{AEDM} is but one of a number of topologies that induce a non-vanishing  $ \bar{g}_{\pi NN}^{(0)} $. A detailed discussion of this and other graphs is given in appendix \ref{AppA}. Since a subset of these diagrams are divergent, one requires a counterterm whose {\em a priori} finite part is analytic in $m_q\sim m_\pi^2$  and $m_\varphi$ and whose value we  estimate to be not larger in magnitude than the calculable loop contributions.
The parts of the latter that are non-analytic  in $m_q$ and $m_\varphi$ cannot be absorbed into the finite part of the counterterm and are, thus, uniquely identified with the loops. For purposes of obtaining our benchmark, order-of-magnitude estimate, it suffices to concentrate on the result for the topology indicated in the third panel of Fig.~\ref{AEDM}, which yields a finite result
\bea
\label{eq:gpiNNshift1} 
\delta \bar g_{\pi NN}^{(0)} = \frac{1}{16\pi}\frac{m_\pi^2 + m_\pi m_\varphi + m_\varphi^2}{m_\pi + m_\varphi}\frac{g_s^\pi g_pg_A}{m_Nf_\pi}\ \ \ ,
\eea
where $g_s^\pi$ is the scalar $\varphi\,\pi\pi$ coupling. As we show in appendix \ref{AppC}, one may relate the $\varphi\,\pi\pi$ and $\varphi\, NN$ couplings as 
\bea
g_s^\pi \simeq\frac{m_\pi^2}{90\>\rm{MeV}}\>g_s\ \ \ ,
\eea
so that the contribution to $ \bar{g}_{\pi NN}^{(0)} $ can be expressed in terms of the product $g_s g_p$. 

One can obtain a conservative bound  (assuming no spurious cancellations with other contributions) on $g_s g_p$ by requiring that the contribution to $d_{Hg}$ {\em via} Eqs.~({\ref{schiff},\ref{dHg}) is less than the current EDM bound given in Eq.~(\ref{EDMlimits}). Using the best value for $a_0$ we then conclude that 
$|g_s g_p| \lesssim 10^{-9}$. As indicated earlier, we na\"ively expect the contribution from the direct $\varphi$-exchange  to be about two orders of magnitude larger. Erring on the conservative side, we thus arrive at a range of upper bounds on $g_s g_p$ lying in the range
\bea
\label{eq:edmrange}
|g_s g_p| \lesssim \big[10^{-11},10^{-9}\big]\ \ \ .
\eea

\section{Comparison of fifth-force and EDM limits}
\label{compare}

\OMIT{
In this section, we discuss and summarize limits on macroscopic SD forces from laboratory fifth-force experiments, EDM constraints, and the analysis of the previous sections. In particular, we discuss the limits on the product of nucleon level couplings $g_sg_p$ of the light scalar mediator $\varphi$, that determine the size of the macroscopic couplings $g_s^1g_p^2$ that determine the fifth-force potential in Eq.~(\ref{potential}). We also discuss and summarize the differences between the case of the axion and the more generic light scalars.
}
\OMIT{
As shown in section  \ref{EDMscalar}, for a generic scalar, the current constraint on the mercury EDM $d_{Hg}$, given in Eq.~(\ref{EDMlimits}), gives an upper bound on $g_sg_p$ in the range
\bea
g_s g_p \lesssim [10^{-11},10^{-9}]\ \ \ .
\eea
}

The bound in Eq.~(\ref{eq:edmrange}) can be compared with those arising from laboratory fifth-force experiments. From Fig.~3 of Ref.~\cite{Tullney:2013wqa} the bound on $g_sg_p$ for two different interaction ranges are given in Table~\ref{SDforce}.
\begin{table}
\renewcommand{\arraystretch}{1.5}
\begin{center}
\begin{tabular}{|c|c|c|c|c|}
\hline
Range & Fifth Force & EDM &  EDM & Combined Laboratory \\[-1.5ex]
$\lambda$ [m]  & (Axion or Generic Scalar) & (Generic Scalar)  & (Axion) &  \& Astrophysics \\[-1.5ex]
&&&& (Axion or Generic Scalar) \\
\hline\hline
$\sim 2 \times 10^{-5}$ & $\sim 10^{-16}$ & $ \sim 10^{-9}-10^{-11}$ & $\sim 10^{-33}$& $\sim 10^{-27}$ \\
\hline
$\sim 2 \times 10^{-1}$ & $\sim 10^{-29}$ & $ \sim 10^{-9}-10^{-11}$ & $\sim 10^{-41} $ & $\sim10^{-30}-10^{-34}$\\
\hline
\end{tabular}
\end{center}
\caption{Comparison of the upper bound on $g_sg_p$ from fifth-force and EDM experiments and from combining astrophysical limits with laboratory constraints.  For the special case of the axion, the EDM limit dominates. For a generic scalar, fifth-force and combined laboratory  limits dominate for the range of interactions they probe. Thus, the relative strength of EDM and laboratory/astrophysics limits depends strongly on whether the underlying force-mediator is an axion or a generic scalar.}
\label{SDforce}
\end{table}
In this case, one can conclude that the laboratory fifth-force experiments place more stringent bounds by several orders of magnitude. Also note that the bounds from laboratory fifth-force experiments exhibit far greater sensitivity to the interaction range, or equivalently to the mass $m_{\varphi}$. This is simply understood by noting that EDM constrains have no sensitivity to $m_\varphi$, since the typical nuclear size $r_N \ll 1/m_\varphi$; in short, compared to typical nuclear scales, the light scalar $\varphi$ is essentially massless. Only when $1/m_\varphi \sim r_N \sim 1/m_\pi$ can one expect EDM bounds to be sensitive to $m_\varphi$   (see {\em e.g.}  Eq.~(\ref{eq:gpiNNshift1})). From Ref.~\cite{Antoniadis:2011zza}, this may occur somewhere in the region where $10^{-10}$ m $\lesssim\lambda\lesssim10^{-7}$ m, corresponding to $2$ eV $\lesssim m_\varphi\lesssim2$ keV. Finally, for $\lambda \lesssim 10^{-10}$ m, corresponding to $m_\varphi \gtrsim 2$ keV, one expects EDM limits to dominate over those from fifth-force experiments.
However,  in this case the interaction range is too small for it to be observed as a macroscopic SD force.

%The interaction ranges  listed in Table~\ref{SDforce}, fall within the so called axion window -- the presently allowed range of masses for axions. However, as one increases $m_\varphi$ and goes outside the axion window, one expects the sensitivity of fifth-force experiments to decrease with decreasing interaction range $\lambda$. Since the EDM limits are insensitive to $m_\varphi$, as long as $m_\varphi \ll m_\pi$, one expects that eventually the EDM and fifth-force limits might  become comparable. 

For the generic light scalar, even more stringent bounds on the product $g_sg_p$ are derived by combining existing laboratory limits with limits obtained from energy loss in the observed 1987A supernova. The laboratory limits on $g_s$ from tests of NewtonÕs inverse square law \cite{Decca:2007jq,Sushkov:2011zz,Geraci:2008hb,Kapner:2006si,Hoskins:1985tn}, the weak equivalence principle \cite{Smith:1999cr,Schlamminger:2007ht}, and from astrophysical  limits \cite{Grifols:1986fc,Grifols:1988fv,Raffelt:1999tx}  are combined with the SN 1987A limit on the pseudoscalar coupling $g_p$ (see Fig.~3 in Ref. \cite{Raffelt:2012sp}), to obtain the most stringent limits, as seen in the last column of Table \ref{SDforce}. Nevertheless, pure laboratory searches remain important, especially if with improvements over time they can compete with astrophysical limits\footnote{T. G. Walker, private communication}.

For the case of axion-mediated TVPV spin-dependent forces, the situation is reversed. As discussed in section \ref{axionphysics}, the linear dependence of $g_s$ on ${\bar\theta}$ that, in turn, is severly constrained by EDM searches, implies that the fifth-force bounds on $g_s g_p$ are several orders of magnitude weaker (see Eq.~(\ref{eq:axionbound})). Numerically, the EDM constraints on $g_sg_p$ for the axion take the form~\cite{Rosenberg:2000wb}
\bea
\label{axionestimate}
g_s g_p \lesssim \>\theta_{\rm eff} \> \Big [\frac{1 \>{\rm mm}}{\lambda} \Big ]^2 6 \times 10^{-27},
\eea
where  $\lambda$ is the Compton wavelength of the axion obtained in terms of the axion mass which is related to the Peccei-Quinn symmetry breaking scale $m_a \sim 1/f_a$, as seen for the case of one quark flavor in Eq.(\ref{gasp}). Thus, unlike the case of the generic scalar, EDM constraints are sensitive to the axion Compton wavelength since $g_sg_p\propto m_a^2$. More recent ~\cite{Roberts:2011wy} calculations of the quark condensates do not affect the order of magnitude of the estimate in Eq.~(\ref{axionestimate}).
For this axion scenario, the fifth-force searches cannot compete with EDM limits, as seen in Table \ref{SDforce} where we have used the bound $\theta_{\rm eff} <10^{-10}$.

%it is well-known that for axions the EDM limits dominate over fifth-force limits by several orders of magnitude. This is due to the fact that couplings of the axions to SM quarks are related to the strong CP parameter and in this case the resulting bound is (see section \ref{axionphysics}) 
%\bea
%\label{EDMaxion}
%g_s g_p\big |_{\text{axion}} \propto \theta_{\text{eff}} \frac{m_q^2}{f_a^2} \lesssim [10^{-40},10^{-34}],
%\eea
%where the constant of proportionality is determined by the relevant nucleon matrix elements.
%Thus, for the case of axions, fifth-force experiments currently cannot compete with EDM limits.

Finally, we  note that the dependence of the nucleon level couplings $g_s, g_p$ on $m_\varphi$ is different for the axion compared to a more generic scalar.  In the case of the axion, the mass is $m_a \sim 1/f_a$ (see Eq.~\ref{gasp}), so we have $g_{s} g_{p} \sim 1/f_a^2 \sim m_a^2$, as seen in Eq.~(\ref{gs1gp2a}). On the other hand, as already discussed, for the case of a generic scalar the nucleon level couplings $g_s,g_p$ are  independent of the mass $m_\varphi$.  Thus, while EDM constraints are largely insensitive to the light scalar mass $m_\varphi$ in the case of generic scalars, they do exhibit sensitivity for the special case of the axion.

%%%%%%%%%%%%%%%%%%%%%%%%%%%%%%
\section{Conclusion}  
\label{sec:conclude}

If a non-zero signal is observed in EDM and/or fifth-force experiments,  and if the culprit is an interaction mediated by the exchange of an ultra-light spin-zero particle, a comparison of results from the two classes of laboratory experiments considered here -- along with the indirect astrophysical constraints -- could provide insight into the nature of the new boson. If, for example, an EDM signal is observed with no corresponding signal in fifth-force experiments, then consistency with the astrophysical bounds would suggest that either the new particle is an axion or that the range is microscopic rather than macroscopic. 
On the other hand, observation of a non-zero spin-dependent TVPV effect in fifth-force experiments with no corresponding EDM signal would point to a generic (non-axion) light scalar. Consistency with the astrophysical bounds would then indicate a range that is order tens of centimeters or larger. Finally, the observation of non-zero signals in both classes of experiments would again point to the generic light scalar mediating the fifth force signal, while an alternate mechanism would likely be responsible for a non-vanishing EDM. Any of these outcomes would constitute a remarkable discovery, and its pursuit is well worth the effort on all fronts.

%In this paper, we have explored the possibility of correlating EDM constraints and laboratory fifth-force experiments as a way to constrain new macroscopic spin-dependent (SD) forces. It is well-known that for the case of SD forces mediated by the axion, invoked to solve the strong CP problem, EDM constraints give the most stringent bounds. However, we have shown that for more generic scalars, unrelated to the strong CP problem, the EDM and fifth-force constraints can be competitive and that  either one can dominate over the other depending on the interaction range of the new SD force. For interactions on the macroscopic scale, the fifth-force experiments yield the most stringent bounds, while for forces that act on much smaller scales, EDM searches provide the most sensitive laboratory probes.

%We advocate the use of EDM and laboratory fifth-force experiments in a complementary manner to probe the nature of any new CP violation or SD forces that may lie beyond the Standard Model. 

\subsection*{Acknowledgements}
   
We acknowledge fruitful discussions with H. Abele, P. Chu, H. Gao, and T. G. Walker. 
This work was supported in part by: U. S. Department of Energy contracts  DE-AC02-06CH11357 (MP), DE-FG02-08ER41531 (MP and MJRM), and DE-SC0011095 (MJRM), the Wisconsin Alumni Research Foundation (MP and MJRM), Northwestern University (SM), and the theoretical program on the contract I689-N16 by the Austrian ÒFonds zur F\"orderung der Wissenschaftlichen ForschungÓ (MP).

\appendix

%%%%%%%%%%%%%%
%%%%%%%%%%%%%%
%%%%%%%%%%%%%%
\section{Computation of the TVPV one-loop diagrams}\label{AppA}

In this section, we outline the calculation leading to the shift in the pion-nucleon couplings shown in Eq.~(\ref{eq:gpiNNshift1}). These shifts arise from the two diagrams shown in Fig.~\ref{HBChPT}, one of which was shown in Fig.~\ref{AEDM}.  We employ techniques of Heavy Baryon Chiral Perturbation Theory (HB$\chi$PT) \cite{Jenkins:1992pi} for the computation.

In addition to the diagrams in Fig.~\ref{HBChPT}, there are additional one-loop diagrams that can contribute to the shift in the pion-nucleon couplings. These diagrams are either suppressed according to the power counting in HB$\chi$PT or involve $\varphi\pi NN$ and $\varphi\varphi\pi NN$ couplings. For completeness, we discuss these diagrams in appendix  
 \ref{appB}. However, the goal here is to give an order of magnitude estimate of the contribution to $d_{Hg}$ of the nucleon-nucleon potential from the tree-level exchange in Fig.~\ref{AEDM}. For this purpose, it is sufficient to estimate it as being enhanced by $16\pi^2$ relative to the contributions from the diagrams in Fig.~\ref{HBChPT}. The additional one-loop diagrams are not expected to change this order of magnitude estimate.

\subsection{Leading contributions}

The shift in the pion-nucleon coupling in Eq.~(\ref{eq:gpiNNshift1}) arises from the calculation of the two diagrams in Fig.~\ref{HBChPT}.
\begin{figure}[h!]
  \centering
  \vspace{1mm}
  \includegraphics[width=0.75\linewidth]{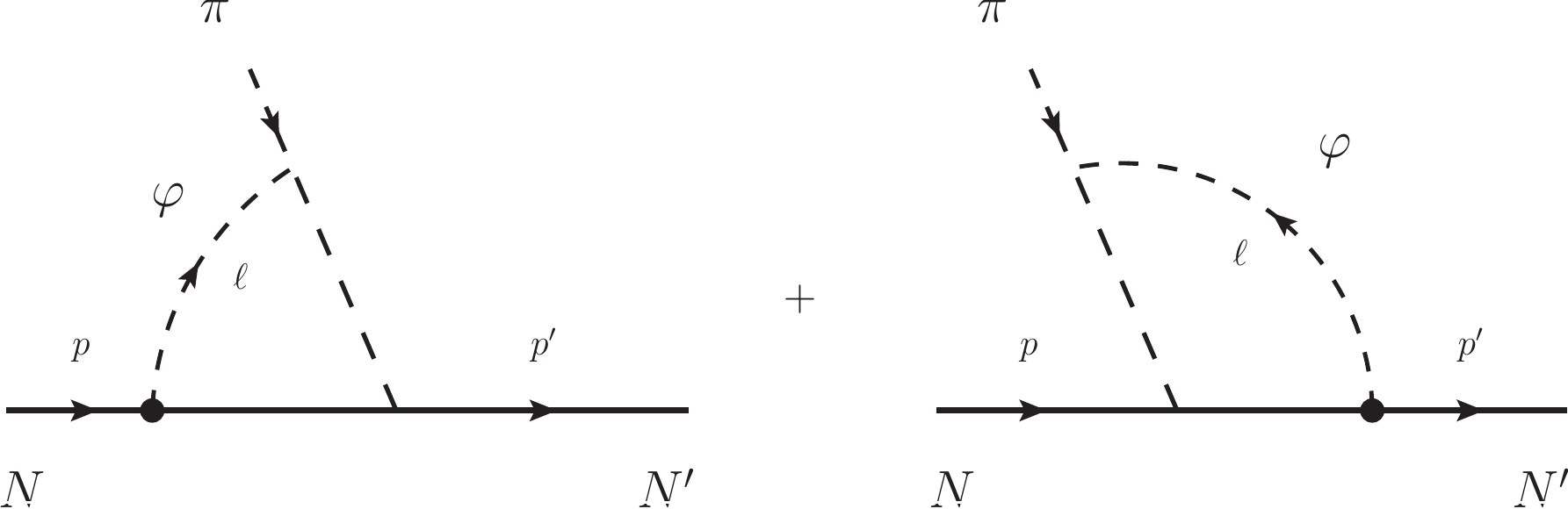}
  \caption{Leading contributions from a virtual $\varphi$ loop that give rise to the shift in the CP-odd pion-nucleon coupling in Eq.~(\ref{eq:gpiNNshift1}).}
  \label{HBChPT}
\end{figure} 
The different vertices in the diagrams are described by the effective interactions in HB$\chi$PT \begin{align}
  \mathcal{L}_{\pi\bar NN}&=\frac{2g_A}{f_\pi}\>\partial_\mu\pi^a\>\bar N_v\>\frac{\tau^a}{2}S^\mu N_v\ \ \ , \\
  \mathcal{L}_{\varphi\pi\pi}&=g_s^\pi\>\varphi\>\pi^a\pi^a\ \ \ , \label{PSC}\\
  \mathcal{L}_{\varphi\bar NN}&=-\frac{g_p}{m_N}\>\bar N_v\left(S^\mu\partial_\mu\varphi\right)N_v\ \ \ , 
\end{align}
where $g_A\simeq1.27$, $m_N\simeq940$ MeV denotes the nucleon mass, and  $f_\pi\simeq 92.4$ MeV is the pion decay constant. The heavy baryon nucleon fields $N_v$ are defined in terms of the full theory nucleon fields $N$ as
\begin{align}
  N_v(x)&=\exp\!\left(im_N\>v\cdot x\right)\frac{1 + \slashed v}{2}\>N(x)\ \ \ ,
\end{align}
where $v^\mu$ denotes the four-velocity which satisfies $v^2=1$. The spin operator $S_\mu$ appearing in Eq.~(\ref{PSC}) is given by
\bea
\label{spinop}
S_\mu &=& \frac{i}{2}\> \gamma^5 \sigma_{\mu \nu} v^\nu\ \ \ , 
\eea
and obeys the relations 
\begin{align}\label{spop}
  S\cdot v=0\>, \qquad [S_\mu,S_\nu] = i\varepsilon_{\mu\nu\alpha\beta} v^\alpha S^\beta\ \ \ .
\end{align}
In appendix~\ref{AppC}, it is shown that the coupling $g_s^\pi$, appearing in $  \mathcal{L}_{\varphi\pi\pi}$ in Eq.~(\ref{PSC}), can be written as $g_s^\pi \simeq\frac{m_\pi^2}{90\>\rm{MeV}}\>g_s$, so that both diagrams are proportional to $g_sg_p$.

The amplitude of the first diagram in Fig.~(\ref{HBChPT}) is given by 
\begin{align}\label{ML}
  i\mathcal{M}_1^a=\frac{g_s^\pi g_pg_A}{m_Nf_\pi}\int\frac{d^d\ell}{(2\pi)^d}\>\bar N_v(p')\tau^a\left(S\cdot\bar q\right)\left(S\cdot\ell\right)N_v(p)\>\frac{1}{v\cdot\bar p + i\varepsilon}\frac{1}{\ell^2 - m_\varphi^2 + i\varepsilon}\frac{1}{\bar q^2 - m_\pi^2 + i\varepsilon}\ \ \ ,
\end{align}
with $q=p' - p$, $\bar p=p - \ell$, $\bar p'=p' + \ell$ and $\bar q=q + \ell$ and  $\bar N_v(p)$ denotes the nucleon $SU(2)$ isospinor in momentum space. The superscript `$a$' on the amplitude denotes the pion isospin index. The amplitude for the second diagram is given by  
\begin{align}\label{MR}
  i\mathcal{M}_2^a=\frac{g_s^\pi g_pg_A}{m_Nf_\pi}\int\frac{d^d\ell}{(2\pi)^d}\>\bar N_v(p')\tau^a\left(S\cdot\ell\right)\left(S\cdot\bar q\right)N_v(p)\>\frac{1}{v\cdot\bar p' + i\varepsilon}\frac{1}{\ell^2 - m_\varphi^2 + i\varepsilon}\frac{1}{\bar q^2 - m_\pi^2 + i\varepsilon}\ \ \ .
\end{align}
Both integrals can be solved exactly. Since the long-range limit $q^\mu \to 0$ of the integral provides a good approximation, we solve the integrals in this limit and the resulting expressions are more compact. Details of the calculation can be found in appendix \ref{AppB}. 
The result of computing the sum of these two diagrams is
\begin{align}
  i\mathcal{M}^a=\frac{i}{16\pi}\frac{m_\pi^2 + m_\pi m_\varphi + m_\varphi^2}{m_\pi + m_\varphi}\frac{g_s^\pi g_pg_A}{m_Nf_\pi}\>\bar N_v(p')\tau^aN_v(p)\ \ \ ,
\end{align}
which we recast as the effective interaction
\begin{align}
\label{CPVPNN}
  \mathcal{L}_{\pi\bar NN}^{CPV}=\frac{1}{16\pi}\frac{m_\pi^2 + m_\pi m_\varphi + m_\varphi^2}{m_\pi + m_\varphi}\frac{g_s^\pi g_pg_A}{m_Nf_\pi}\>\pi^a\bar N\>\tau^a N\ \ \ ,
\end{align}
to be interpreted as a correction to the TVPV pion-nucleon coupling
\bea
\label{eq:gpiNNshift} 
\delta \bar g_{\pi NN}^{(0)} = \frac{1}{16\pi}\frac{m_\pi^2 + m_\pi m_\varphi + m_\varphi^2}{m_\pi + m_\varphi}\frac{g_s^\pi g_pg_A}{m_Nf_\pi}\ \ \ .
\eea

After using $g_s^\pi \simeq\frac{m_\pi^2}{90\>\rm{MeV}}\>g_s$, the shift in the TVPV pion-nucleon couplings is given by Eq.~(\ref{eq:gpiNNshift1}).

\subsection{Additional one-loop diagrams}
\label{appB}

Here we discuss additional one loop diagrams that are either subleading or diagrams generated from higher dimensional vertices. 

\subsubsection{Sub-leading contributions}

Additional contributions arise from diagrams with the scalar having both couplings to the nucleons as shown in Fig.~\ref{nEDM2}.  
\begin{figure}[h!]
  \centering
  \vspace{1mm}
  \includegraphics[width=0.6\linewidth]{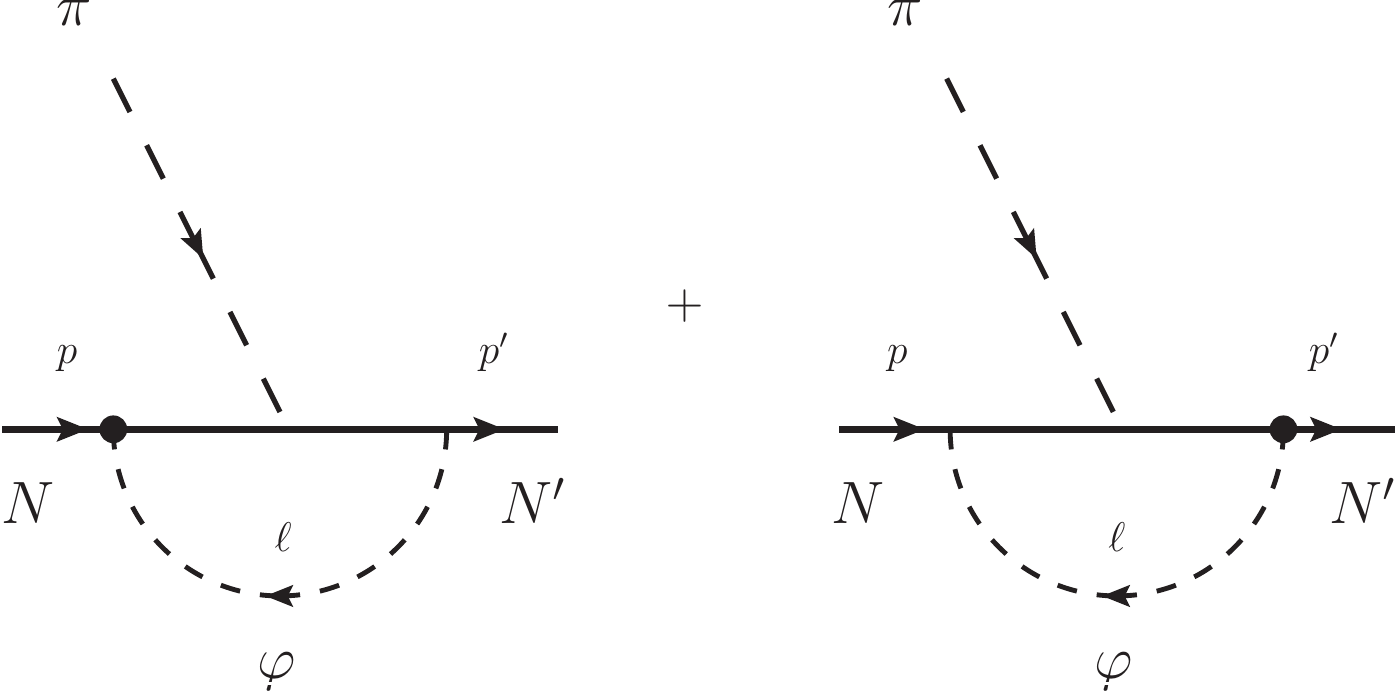}
  \caption{The diagrams with the scalar coupling to nucleons. Each blob depicts the pseudo-scalar coupling.}
  \label{nEDM2}
\end{figure} 
To leading order in $1/m_N$ expansion, the loop integral for the left diagram in Fig.~\ref{nEDM2} is
\begin{align} 
  \int\frac{d^d\ell}{(2\pi)^d}\>\frac{S\cdot \ell}{v\cdot(p + \ell) + i\varepsilon}\frac{1}{v\cdot(p' + \ell) + i\varepsilon}\frac{1}{\ell^2 - m_\varphi^2 + i\varepsilon}= S_\mu\>\mathcal I^\mu(v, v\cdot p, v\cdot p', m_\varphi^2)\ \ \ ,
\end{align}
where $S^\mu$ is the HB$\chi$PT spin operator of Eq.~(\ref{spinop}) and the factor $S\cdot \ell$ in the numerator is due to the derivative pseudo-scalar coupling of scalar to nucleon shown in  Eq.~(\ref{PSC}). Since $v^\mu$ is the only four-vector that the integration variable $\ell^\mu$ is contracted with in the integrand, the vector quantity $\mathcal I^\mu$ must be proportional to $v^\mu$ so that
\begin{align}
  \mathcal I^\mu(v, v\cdot p, v\cdot p', m_\varphi^2) =  \mathcal J(v, v\cdot p, v\cdot p', m_\varphi^2) \> v^\mu\ \ \ ,
\end{align}
where $\mathcal J(v, v\cdot p, v\cdot p', m_\varphi^2)$ is a scalar integral. Therefore, 
\begin{align} 
  \int\frac{d^d\ell}{(2\pi)^d}\>\frac{S\cdot \ell}{v\cdot(p + \ell) + i\varepsilon}\frac{1}{v\cdot(p' + \ell) + i\varepsilon}\frac{1}{\ell^2 - m_\varphi^2 + i\varepsilon}\propto S\cdot v = 0\ \ \ ,
\end{align}
as dictated by the properties of the spin operator $S_\mu$ shown in Eq.~(\ref{spop}). For the same reason, the diagram on the right in Fig.~\ref{nEDM2} also vanishes. Thus, to leading order in $1/m_N$, the two diagrams in Fig.~\ref{nEDM2} give vanishing contributions. 

Next we consider the TVPV nucleon wave-function renormalization diagrams, proportional to $g_sg_p$.  Two of these are depicted in Fig.~\ref{WFR},
\begin{figure}[h!]
  \centering
  \vspace{1mm}
  \includegraphics[width=0.8\linewidth]{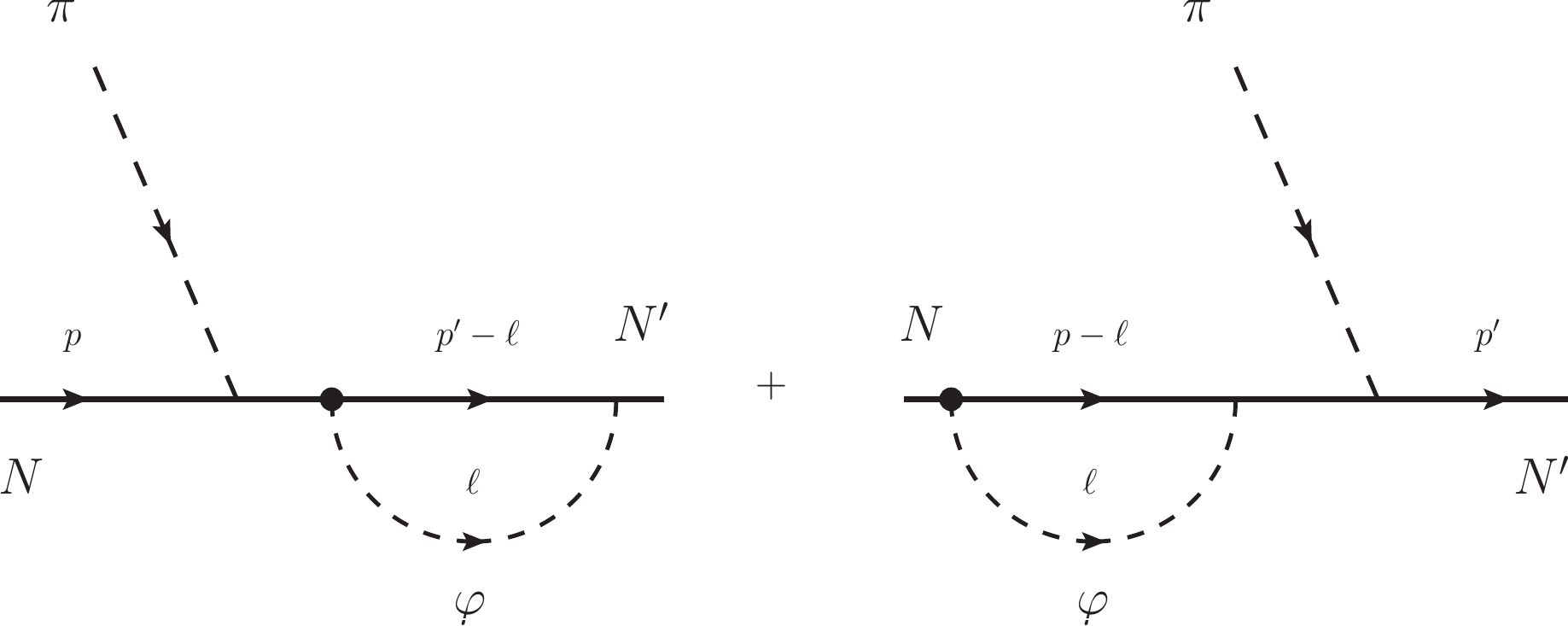}
  \caption{Two of the four wave-function renormalization diagrams. Each dark blob depicts the pseudo-scalar coupling $g_p$, as defined in Eq.~(\ref{axionnucleon}).}
  \label{WFR}
\end{figure} 
and the remaining two are identical except for the interchange of the scalar and pseudo-scalar (dark blob) couplings.  To leading order in the $1/m_N$ expansion, the loop integral for the left diagram in Fig.~\ref{WFR} gives
\begin{align}
  \int\frac{d^d\ell}{(2\pi)^d}\>\frac{S\cdot\ell}{v\cdot(p' - \ell) + i\varepsilon}\frac{1}{\ell^2 - m_\varphi^2 + i\varepsilon}\ \ \ .
\end{align}
Once again,  this integral vanishes since it must be proportional to $S\cdot v=0$. Similarly, all the other TVPV nucleon wave function renormalization diagrams vanish and do not contribute to the EDM at leading order in $1/m_N$.  Note that since the pion has no pseudoscalar coupling to $\varphi$, its wave function diagrams are not proportional to the product $g_sg_p$ and thus also do not contribute to the EDM.  

\subsubsection{Proton EDM}\label{PEDM}

Here we comment  on the $\varphi$ loop-induced contribution to the proton EDM (Fig.~\ref{AEDM}, middle panel). Diagrammatically, the situation is similar to Fig.~\ref{nEDM2} but with the external pion replaced by a photon. The Lagrangian for the nucleon-photon coupling in HBChPT is given by
\begin{align}
  \mathcal{L}_{A\bar NN} = e\>v_\mu A^\mu\>\bar N_v\>\frac{1 + \sigma^3}{2}\>N_v\ \ \ .
\end{align}
Following the same procedure as for the computation of Fig.~\ref{nEDM2}, with the pion vertex replaced by the above photon coupling, to leading order in $1/m_N$ expansion, the same loop integrals as in diagram Fig.~\ref{nEDM2} appear and give vanishing contributions
\begin{align} 
  S_\mu\>\mathcal I^\mu(v, v\cdot p, v\cdot p', m_\varphi^2)\propto S\cdot v = 0\ \ \ .
\end{align}
Thus, the $\varphi$ loop-induced proton EDM vanishes to leading order in $1/m_N$. 

\subsubsection{Diagrams with four or five point vertices}

Here we consider the remaining one-loop diagrams shown in Fig.~\ref{4v1}. 
These contributions have a more complicated structure compared to the diagrams in Fig.~\ref{HBChPT}. In addition to the dependence on the product of couplings $g_sg_p$ of interest, the diagrams in Fig.~\ref{HBChPT} depend on a non-perturbative matrix element through the coupling $g_s^\pi$, as explained in appendix \ref{AppC}. The diagrams in Fig.~\ref{4v1}, however,  depend on new types of non-perturbative matrix elements. Furthermore, unlike the diagrams in Fig.~\ref{HBChPT}, these contributions involve ultraviolet divergences and depend on the renormalization scheme. For the sake of completeness,  we discuss these contributions and where appropriate we give results for the finite non-analytic parts of the contribution that cannot be removed via the renormalization counterterm. 

\begin{figure}[h!]
  \centering
  \vspace{1mm}
  \includegraphics[width=0.9\linewidth]{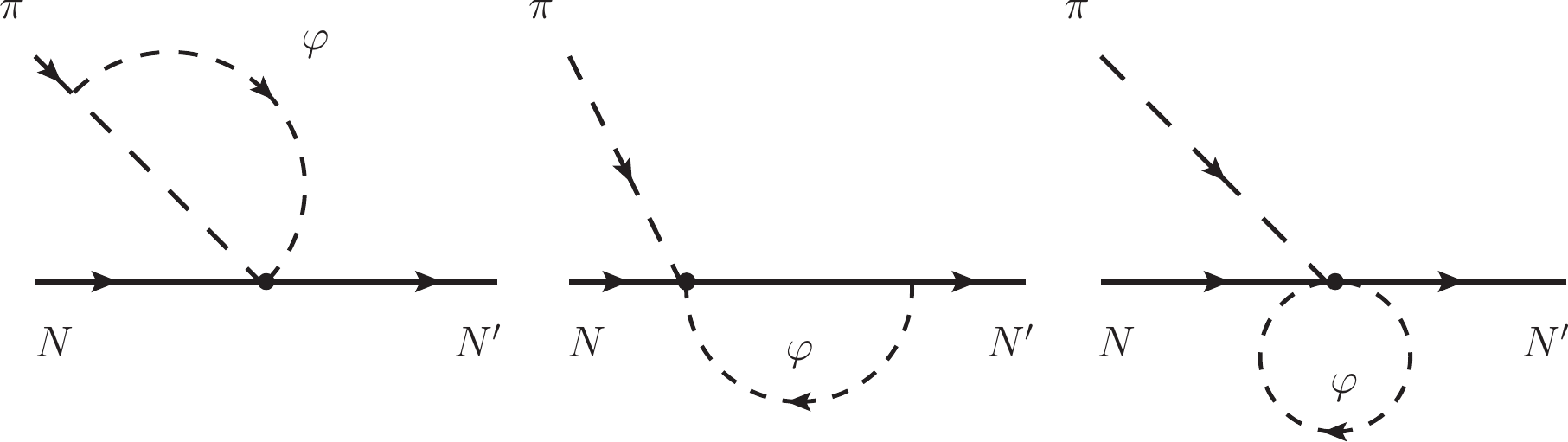}
  \caption{Diagrams with $NN\pi\varphi$ and $NN\pi\varphi^2$ vertices.}
  \label{4v1}
\end{figure}

The vertex in the first diagram in Fig.~\ref{4v1}, denoted by $\lambda_p$, is determined in terms of the quark level coupling $g_p^q$ in Eq.~(\ref{axionquark}) through the matching equation 
\bea
\label{lp1}
 \lambda_p\> \langle \varphi \pi^0 N|\varphi\> \pi^a \bar{N}\tau^a N | N\rangle&=&ig_p^q\>\langle \varphi \pi^0 N| \varphi\>\bar{q}\gamma^5 q | N\rangle\ \ \ .
 \eea
Using the soft pion theorem relation
\bea
{\rm lim}_{k^\mu \to 0}\langle \pi^0 (k) N| \bar{q}\gamma^5 q | N\rangle = -\frac{i}{f_\pi}\>\langle N| \big [Q_5^3,\bar{q}\gamma^5 q\big ] |N\rangle =\frac{i}{f_\pi}\>\langle N|\bar{q}\tau^3q|N\rangle\ \ \ ,
\eea
Eq.~(\ref{lp1}) then leads to 
\bea
\label{lp1b}
 \lambda_p\>  \bar{N}\tau^3 N &=& -\frac{g_p^q}{f_\pi}\>\langle N | \bar{q} \tau^3 q |N\rangle\ \ \ .
 \eea
The matrix element on the right hand side of Eq.~(\ref{lp1b}) can be related to the light quark contribution to the neutron-proton mass difference, $(\Delta m_N)_q$
\bea
\langle N | \bar{q} \tau^3 q |N\rangle  = \frac{(\Delta m_N)_q}{m_d-m_u}\> \bar{N} \tau^3 N\ \ \ ,
\eea
so that 
\bea
\lambda_p = -\frac{g_p^q}{f_\pi}\frac{(\Delta m_N)_q}{m_d-m_u}\ \ \ .
\label{lp1c}
\eea
We now wish to relate $g_p^q$ to the effective $\varphi\, NN$ pseudoscalar coupling
\bea
g_p^q\>\langle\varphi N| \varphi\> \bar{q} i\gamma^5 q | N \rangle = g_p\> \langle \varphi N|  \varphi\> \bar{N} i\gamma^5 N | N \rangle\ \ \ .
\eea
Using 
\bea
\langle\varphi N| \varphi\> \bar{q} i\gamma^5 q | N \rangle = \langle N| \bar{q} i\gamma^5 q | N \rangle = 2 G_P^{(0)} \bar{N} i\gamma^5 N\ \ \ ,
\eea
where $G_P^{(0)}$ is the isoscalar nucleon pseudoscalar form factor at zero momentum transfer, we have
\bea
g_p= 2G_P^{(0)}g_p^q\ \ \ .
\label{gpq}
\eea
Substituting this result into Eq.~(\ref{lp1c}) leads to
\bea
\label{lp1d}
\lambda_p = -\frac{g_p}{2G_P^{(0)} f_\pi}\> \frac{(\Delta m_N)_q}{m_d-m_u}\ \ \ .
\eea

With the result in Eq.~(\ref{lp1d}) and the corresponding relation between $g_s$ and the induced scalar coupling $g_s^\pi$ given in Eq.~(\ref{gs-gspi}) below, we are able to compute the contribution to $ \delta \bar g_{\pi NN}^{(0)}$ arising from the first diagram of Fig.~\ref{4v1}. The graph itself is divergent, thereby implying the need for a counterterm and an associated finite part that must be analytic in $m_\pi$ and $m_\varphi$. The contribution uniquely associated with the loop is non-analytic in these masses and is given by 
\bea
\label{eq:loop4v1a}
\delta \bar g_{\pi NN}^{(0)} = - \frac{g_s^\pi\lambda_p}{16\pi^2}\left(\frac{m_\pi^2\log (m_\pi^2/\mu^2)-m_\varphi^2\log (m_\varphi^2/\mu^2)}{m_\pi^2 -m_\varphi^2}\right)
\ \ \ .
\eea

We now turn to the second diagram of Fig.~\ref{4v1}, wherein the $\varphi$ couples to the nucleon through the pseudoscalar interaction and to the $NN\pi$ through the scalar interaction. To evaluate the latter vertex, we follow a similar logic to that of the foregoing computation, starting with the matching equation
\bea
\lambda_s \> \langle \varphi \pi^0 N| \varphi\>\pi^a \bar{N}\gamma^5 \tau^a N | N\rangle &=& i g_s^q\> \langle \pi^0 N|\varphi\> \bar{q}q | N\rangle\ \ \ ,
\eea
and the soft pion relation
\bea
{\rm lim}_{k^\mu \to 0}\langle \pi^0(k) N| \bar{q} q | N\rangle = \frac{i}{f_\pi}\>\langle N|\bar{q}\gamma^5\tau^3q|N\rangle\ \ \ , 
\eea
to obtain
\bea
\label{eq:ls1}
\lambda_s\> \bar N\gamma^5\tau^3 N = -\frac{g_s^q}{f_\pi}\> \langle N|\bar{q}\gamma^5\tau^3q|N\rangle\ \ \ .
\eea
The latter matrix element is given by 
\bea
\label{eq:ls2}
\langle N|\bar{q}\gamma^5\tau^3q|N\rangle = 2 G_P^{(1)} \bar{N}\gamma^5\tau^3 N\ \ \ ,
\eea
where the isovector pseudoscalar form factor at zero momentum transfer is given by 
\bea
\label{eq:ls3}
G_P^{(1)} = \frac{2 g_A \bar{m}_N}{m_u+m_d}\ \ \ ,
\eea
with $\bar{m}_N$ being the average of the neutron and proton masses. The coupling $g_s^q$ can be related to the effective $\varphi\, NN$ scalar coupling through
\bea
g_s\>\langle \varphi N| \varphi\> \bar{N}N| N\rangle = g_s^q\>\langle \varphi N| \varphi\> \bar{q} q | N\rangle\ \ \ , 
\eea
or
\bea
\label{eq:ls4}
g_s\> \bar{N} N = g_s^q\>\langle N| \bar{q} q | N\rangle = 2 g_s^qG_S^{(0)} \bar{N} N\ \ \ ,
\eea
with $G_S^{(0)} $ being the iscoscalar nucleon scalar form factor at zero momentum transfer given in terms of the light quark contribution to the average nucleon mass  $(\bar{m}_N)_q$ by
\bea
G_S^{(0)} = \frac{(\bar{m}_N)_q}{m_u+m_d}\ \ \ .
\eea 
Using Eqs.~(\ref{eq:ls1}) and (\ref{eq:ls2}) in Eq.~(\ref{eq:ls4}) we obtain
\bea
\label{eq:ls5}
\lambda_s = - \frac{g_s}{f_\pi}\frac{G_P^{(1)}}{G_S^{(0)}}\ \ \ .
\eea
The second loop in Fig.~\ref{4v1} is finite and gives
\bea
\label{eq:loop4v1b}
\delta \bar g_{\pi NN}^{(0)} = - \frac{g_p\lambda_s}{8\pi}\> m_\varphi\ \ \ .
\eea

The five-point vertex appearing in the last diagram in Fig.~\ref{4v1} contains the time ordered product of operators $g_s^q\, \bar{q}q$ and $g_p^q\, \bar{q}i\gamma^5 q$. Its evaluation is non-trivial and goes beyond the scope of the present study whose aim is to provide order of magnitude estimates. Consequently, we now restrict our attention to the results for the first two graphs of Fig.~\ref{4v1}. We wish to compare the magnitudes of the induced shifts in Eqs.~(\ref{eq:loop4v1a},\ref{eq:loop4v1b}) to the result obtained from Fig.~\ref{HBChPT} given in Eq.~(\ref{eq:gpiNNshift}). Since (\ref{eq:loop4v1b}) vanishes in the $m_\varphi\to 0$ limit while (\ref{eq:gpiNNshift}) remains finite, the latter will dominate in the regime $m_\pi>>m_\varphi$ that is of interest to the experimental probes of macroscopic P- and T-odd interactions. 
Comparison of (\ref{eq:gpiNNshift}) with (\ref{eq:loop4v1a}) requires choice of a renormalization scale and knowledge of the scalar nucleon isoscalar form factor. 
We will assume that the finite part of the counterterm is of the same magnitude as the loop contribution. 
Working in the $m_\varphi\to 0$ limit we have the ratio $\mathcal{R}$ of (\ref{eq:loop4v1a})  to (\ref{eq:gpiNNshift})  is given by
\bea
\label{eq:R}
\mathcal{R}\simeq \frac{1}{2\pi} \frac{1}{g_A G_P^{(0)}}\frac{m_N}{m_\pi}\frac{(\Delta m_N)_q}{m_d-m_u}\> \ln\frac{m_\pi^2}{\mu^2}\ \ \ .
\eea
Lattice results for $(\Delta m_N)_q$ imply that the fourth factor in Eq.~(\ref{eq:R}) is order one, as is $m_N/(2\pi m_\pi)$. For $\mu\sim 1$ GeV, the ratio $\mathcal{R}$ will then be $\mathcal{O}(1)$ to the extent that $G_P^{(0)}$ is as well. Thus, we conclude that the result in Eq.~(\ref{eq:gpiNNshift}) provides a reasonable, order of magnitude estimate for the loop-induced shifts $\delta \bar g_{\pi NN}^{(0)}$.

\section{Calculational details}\label{AppB}

Here we give details of the  computation of the integrals in Eqs.~(\ref{ML}) and (\ref{MR})
\begin{align}
\label{i1}
  \mathcal{I}_1=\int\frac{d^d\ell}{(2\pi)^d}\left(S\cdot[q + \ell]\right)\left(S\cdot\ell\right)\frac{1}{v\cdot(p - \ell) + i\varepsilon}\frac{1}{\ell^2 - m_\varphi^2 + i\varepsilon}\frac{1}{(q + \ell)^2 - m_\pi^2 + i\varepsilon}\ \ \ ,
\end{align}
and 
\begin{align}
\label{i2}
  \mathcal{I}_2=\int\frac{d^d\ell}{(2\pi)^d}\left(S\cdot\ell\right)\left(S\cdot[q + \ell]\right)\frac{1}{v\cdot(p' + \ell) + i\varepsilon}\frac{1}{\ell^2 - m_\varphi^2 + i\varepsilon}\frac{1}{(q + \ell)^2 - m_\pi^2 + i\varepsilon}\ \ \ ,
\end{align}
respectively.
The nucleon momenta are given by $m_Nv + p$ with residual momentum $p$. Since the typical virtuality of the nucleon inside a nucleus is mach smaller than its mass, we have  $(m_Nv + p)^2\simeq m_N^2$ so that $v\cdot p\simeq-p^2/(2m_N)\ll 1$. For on-shell external nucleons and in the limit $q^\mu\to 0$, we can set $v\cdot p=v\cdot p'=0$ in the computation of the integrals in Eqs.~(\ref{i1}) and (\ref{i2}) so that we get
\begin{align}
  \mathcal{I}_1
  &=\int\frac{d^d\ell}{(2\pi)^d}\left(S\cdot\ell\right)\left(S\cdot\ell\right)\frac{1}{v\cdot\ell + i\varepsilon}\frac{1}{\ell^2 - m_\varphi^2 + i\varepsilon}\frac{1}{\ell^2 - m_\pi^2 + i\varepsilon}\ \ \ , \nn \\
  \mathcal{I}_2&=\int\frac{d^d\ell}{(2\pi)^d}\left(S\cdot\ell\right)\left(S\cdot\ell\right)\frac{1}{v\cdot\ell + i\varepsilon}\frac{1}{\ell^2 - m_\varphi^2 + i\varepsilon}\frac{1}{\ell^2 - m_\pi^2 + i\varepsilon}\ \ \ .
\end{align}
Adding both contributions and using the well-know relation $\displaystyle\{S_\mu,S_\nu\}=\frac{1}{2}\left(v_\mu v_\nu - g_{\mu\nu}\right)$ (see e.g. \cite{Ecker:1993ft}) we obtain
\begin{align}
  \mathcal{I}&=\mathcal{I}_1 + \mathcal{I}_2=-\frac{1}{2}\int\frac{d^d\ell}{(2\pi)^d}\frac{\ell^2}{v\cdot\ell + i\varepsilon}\frac{1}{\ell^2 - m_\varphi^2 + i\varepsilon}\frac{1}{\ell^2 - m_\pi^2 + i\varepsilon}\ \ \ .
\end{align}
Applying the Feynman parametrization we obtain
\begin{align}
  \mathcal{I}=-\frac{1}{2}\int_0^1dx\int\frac{d^d\ell}{(2\pi)^d}\frac{\ell^2}{v\cdot\ell + i\varepsilon}\frac{1}{\left[\ell^2 - x(m_\varphi^2 - m_\pi^2) - m_\pi^2 + i\varepsilon\right]^2}\ \ \ .
\end{align}
Next we use the identity 
\begin{align}
  \frac{1}{a^rb^s}=2^s\frac{\Gamma(r + s)}{\Gamma(r)\Gamma(s)}\int_0^\infty d\lambda\>\frac{\lambda^{s-1}}{(a + 2b\lambda)^{r+s}}\ \ \ ,
\end{align}
to find
\begin{align}
  \mathcal{I}&=-2\int_0^1dx\int_0^\infty d\lambda\int\frac{d^d\ell}{(2\pi)^d}\frac{\ell^2}{\left[\ell^2 - x(m_\varphi^2 - m_\pi^2) - m_\pi^2 + 2\lambda v\cdot\ell+ i\varepsilon\right]^3} \nonumber\\
  &=-2\int_0^1dx\int_0^\infty d\lambda\int\frac{d^d\ell}{(2\pi)^d}\frac{\ell^2 + \lambda^2}{\left[\ell^2 - \lambda^2 - x(m_\varphi^2 - m_\pi^2) - m_\pi^2 + i\varepsilon\right]^3} \nonumber\\
  &\equiv\mathcal{I}_A + \mathcal{I}_B\ \ \ ,
\end{align}
where  $\mathcal{I}_A $ and $\mathcal{I}_B$ correspond to $\ell^2$ and $\lambda^2$ terms in the integrand.
Working in $d=4-2\epsilon$ dimensions, a straightforward computation gives
\begin{align}
  \mathcal{I}_A&=-\frac{i}{(4\pi)^{2 - \epsilon}}\>(2 - \epsilon)\Gamma(\epsilon)\int_0^1dx\int_0^\infty d\lambda\left[\lambda^2 + x(m_\varphi^2 - m_\pi^2) + m_\pi^2 - i\varepsilon\right]^{-\epsilon}\ \ \ , \nonumber\\
  \mathcal{I}_B&=\frac{i}{(4\pi)^{2 - \epsilon}}\>\Gamma(1 + \epsilon)\int_0^1dx\int_0^\infty d\lambda\>\lambda^2\left[\lambda^2 + x(m_\varphi^2 - m_\pi^2) + m_\pi^2 - i\varepsilon\right]^{-1 - \epsilon}\ \ \ .
\end{align}
Next we use the identity 
\begin{align}
  \int_0^1dx\>[Ax + B]^\alpha&=\frac{1}{\alpha + 1}\frac{1}{A}\left\{[A + B]^{\alpha+1} - B^{\alpha+1}\right\}\ \ \ ,
\end{align}
to obtain
\begin{align}
  \mathcal{I}_A&=-\frac{i}{(4\pi)^{2 - \epsilon}}\frac{2 - \epsilon}{1 - \epsilon}\frac{\Gamma(\epsilon)}{m_\varphi^2 - m_\pi^2}\int_0^\infty d\lambda\left\{\left[\lambda^2 + m_\varphi^2 - i\varepsilon\right]^{1-\epsilon} - \left[\lambda^2 + m_\pi^2 - i\varepsilon\right]^{1-\epsilon}\right\}\ \ \ , \nonumber\\
  \mathcal{I}_B&=-\frac{i}{(4\pi)^{2 - \epsilon}}\frac{1}{\epsilon}\frac{\Gamma(1 + \epsilon)}{m_\varphi^2 - m_\pi^2}\int_0^\infty d\lambda\>\lambda^2\left\{\left[\lambda^2 + m_\varphi^2 - i\varepsilon\right]^{-\epsilon} - \left[\lambda^2 + m_\pi^2 - i\varepsilon\right]^{-\epsilon}\right\}\ \ \ .
\end{align}
Next we use the relation
\begin{align}
  \int_0^\infty d\lambda\>\lambda^{2\alpha}[\lambda^2 + m^2]^\beta=\frac{(m^2)^{\alpha + \beta + 1/2}}{2}\int_0^\infty du\>u^{\alpha - 1/2}\>[u + 1]^\beta\ \ \ ,
\end{align}
obtained after the  substitution $\lambda^2=u$ and the definition 
\begin{align}
  B(m,n)=\int_0^\infty du\>\frac{u^{m-1}}{(u + 1)^{m+n}}\ \ \ ,
\end{align}
to get
\begin{align}
  \int_0^\infty d\lambda\>\lambda^{2\alpha}[\lambda^2 + m^2]^\beta=\frac{(m^2)^{\alpha + \beta + 1/2}}{2}\>B\!\left(\alpha+\frac{1}{2},-\alpha - \beta - \frac{1}{2}\right)\ \ \ .
\end{align}
The integrals $\mathcal{I}_A$ and $\mathcal{I}_B$ can now be brought into the form
\begin{align}
  \mathcal{I}_A&=-\frac{i}{(4\pi)^{2 - \epsilon}}\frac{2 - \epsilon}{1 - \epsilon}\frac{\Gamma(\epsilon)}{2}\frac{(m_\varphi^2)^{3/2 - \epsilon} - (m_\pi^2)^{3/2 - \epsilon}}{m_\varphi^2 - m_\pi^2}\>B\!\left(\frac{1}{2},\epsilon - \frac{3}{2}\right)\ \ \ , \nonumber\\
  \mathcal{I}_B&=-\frac{i}{(4\pi)^{2 - \epsilon}}\frac{1}{\epsilon}\frac{\Gamma(1 + \epsilon)}{2}\frac{(m_\varphi^2)^{3/2 - \epsilon} - (m_\pi^2)^{3/2 - \epsilon}}{m_\varphi^2 - m_\pi^2}\>B\!\left(\frac{3}{2},\epsilon - \frac{3}{2}\right)\ \ \ ,
\end{align}
and correspondingly the some of these two terms gives
\begin{align}
  \mathcal{I}=-\frac{i\sqrt\pi}{2}\frac{\Gamma(\epsilon - 3/2)}{(4\pi)^{2 - \epsilon}}\frac{(m_\varphi^2)^{3/2 - \epsilon} - (m_\pi^2)^{3/2 - \epsilon}}{m_\varphi^2 - m_\pi^2}\left\{\frac{2 - \epsilon}{1 - \epsilon}\frac{\Gamma(\epsilon)}{\Gamma(\epsilon - 1)} + \frac{1}{2\epsilon}\frac{\Gamma(1 + \epsilon)}{\Gamma(\epsilon)}\right\}\ \ \ .
\end{align}
Going back to $d=4$ dimensions via the limit $\epsilon\to0$, gives the final result for the sum of the two diagrams in Fig.~\ref{HBChPT} as
\begin{align}
  \mathcal{I}=\frac{i}{16\pi}\frac{m_\varphi^2 + m_\varphi m_\pi + m_\pi^2}{m_\varphi + m_\pi}\ \ \ .
\end{align}

%%%%%%%%%%%%%%%%%%%%%%%%%%%%%%%%%%%%%%%%%%%%%%%%
\section{Scalar coupling to the pion}\label{AppC}

In this section we show that the coupling $g_s^\pi$, appearing in Eq.~(\ref{PSC}), is proportional to the scalar nucleon coupling $g_s$.  We start with the quark level coupling $g_s^q$, assuming  flavor universality for simplicity, so that 
\bea
\label{Lqphi}
  \mathcal{L}_\varphi^q=g_s^q\>\varphi \> \big(\bar uu + \bar dd\big)\ \ \ ,
\eea
which induces a coupling $g_s^\pi$ to pions 
\begin{align}
  \mathcal{L}_\varphi^\pi =g_s^\pi\>\varphi\>\pi^a\pi^a\ \ \ ,
\end{align}
and the coupling $g_s$ to nucleons
\begin{align}
  \mathcal{L}=g_s\>\varphi\>\bar NN\ \ \ .
\end{align}
By taking pion and nucleon matrix elements of the operator ${\cal L}_\varphi^q$ in Eq.~(\ref{Lqphi}),  the quark level coupling $g_s^q$ is related to the pion ($g_s^\pi$) and nucleon ($g_s$) level couplings as
\bea
 g_s^\pi &=& g_s^q\>\langle\pi|\bar uu + \bar dd|\pi\rangle\ \ \ , \nn \\
g_s &=& g_s^q\>\frac{\langle N|\bar uu + \bar dd|N\rangle}{\langle N|\bar NN|N\rangle}\ \ \ ,
\eea
so that $g_s$ and $g_s^\pi$ are related as
\begin{align}
g_s^\pi =\frac{\langle N|\bar NN|N\rangle}{\langle N|\bar uu + \bar dd|N\rangle}\langle\pi|\bar uu + \bar dd|\pi\rangle \>g_s\ \ \ .
\end{align}
We use the relations~\cite{Demir:2003js,Ellis:2008zy} 
\bea
 \frac{\langle N|\bar uu + \bar dd|N\rangle}{\langle N|\bar NN|N\rangle}&\simeq&\frac{90\>\text{MeV}}{m_u + m_d}\ \ \ , \nn \\
 \langle\pi|\bar uu + \bar dd|\pi\rangle&=&\frac{m_\pi^2}{m_u + m_d}\ \ \ ,
\eea
to write\footnote{Note that pion matrix elements of quark bilinears have units of energy in our normalization.}
\bea
\label{gs-gspi}
  \frac{g_s^\pi}{g_s}&\simeq&\frac{m_\pi^2}{90\>\rm{MeV}}\simeq 218\>\rm{MeV}\ \ \ .
\eea

\bibliographystyle{h-physrev3.bst}
\bibliography{fifthforce}

\end{document}